%% file: EccEnv.tex
\documentclass[usenatbib]{mnras}

\usepackage[final]{microtype}
\usepackage[T1]{fontenc}
\usepackage{pifont,stmaryrd} 
\usepackage{ae,aecompl}
\usepackage{graphicx}
\usepackage{amsmath}
\usepackage{amssymb}	
\usepackage{subfig}
\usepackage{multirow}
\usepackage{tabularx}
\usepackage[mathscr]{euscript}
\usepackage{pifont}
\usepackage{array}
\usepackage[usenames,dvipsnames,table]{xcolor}
\usepackage{braket}
\usepackage{enumerate}
\usepackage{amsfonts}

\usepackage{scalefnt}
\usepackage[normalem]{ulem}
\usepackage{tikz}
\usepackage{float}
\restylefloat{table}
\newcolumntype{C}{>{$}c<{$}}
\usepackage{booktabs} 
\newcolumntype{P}[1]{>{\centering\arraybackslash}p{#1}}
\def\simless{\mathbin{\lower 3pt\hbox
{$\rlap{\raise 5pt\hbox{$\char'074$}}\mathchar"7218$}}}   
\def\simmore{\mathbin{\lower 3pt\hbox
{$\rlap{\raise 5pt\hbox{$\char'076$}}\mathchar"7218$}}}   
\newcommand{\be}{\begin{equation}}
\newcommand{\ee}{\end{equation}}
\newcommand{\MSun}{{\rm M}_{\sun}}

\title[Measure eccentricity \& gas from GWs of MBHBs]{Measuring eccentricity and gas-induced perturbation from gravitational waves of LISA massive black hole binaries}
\author[M. Garg et al.]{Mudit Garg,$^{1}$\thanks{E-mail: mudit.garg@uzh.ch} Andrea Derdzinski,$^{1,2,3}$ Shubhanshu Tiwari,$^4$ Jonathan Gair,$^5$ and Lucio Mayer$^1$\\
$^1$Department of Astrophysics, University of Zurich, Winterthurerstrasse 190, CH-8057 Z\"urich, Switzerland\\
$^2${Department of Life and Physical Sciences, Fisk University, 1000 17th Avenue N., Nashville, TN 37208, USA} \\
$^3${Department of Physics \& Astronomy, Vanderbilt University,
2301 Vanderbilt Place, Nashville, TN 37235, USA}\\
$^4$Physik-Institut, Universit\"at Z\"urich, Winterthurerstrasse 190, 8057 Z\"urich, Switzerland\\
$^5${Max Planck Institute for Gravitational Physics (Albert Einstein Institute), Am M\"uhlenberg 1, Potsdam 14476, Germany}}

\date{Received / Accepted}
\pubyear{2024}
\begin{document}
\label{firstpage}
\pagerange{\pageref{firstpage}--\pageref{lastpage}}
\maketitle

\begin{abstract}
We assess the possibility of detecting both eccentricity and gas effects (migration and accretion) in the gravitational wave (GW) signal from LISA massive black hole binaries (MBHBs) at redshift $z=1$. Gas induces a phase correction to the GW signal with an effective amplitude ($C_{\rm g}$) and a semi-major axis dependence (assumed to follow a power-law with slope $n_{\rm g}$). We use a complete model of the LISA response, and employ a gas-corrected post-Newtonian in-spiral-only waveform model \textsc{TaylorF2Ecc}. By using the Fisher formalism and Bayesian inference, we constrain $C_{\rm g}$ together with the initial eccentricity $e_0$, the total redshifted mass $M_z$, the primary-to-secondary mass ratio $q$, the dimensionless spins $\chi_{1,2}$ of both component BHs, and the time of coalescence $t_c$. We find that simultaneously constraining $C_{\rm g}$ and $e_0$ leads to worse constraints on both parameters with respect to when considered individually. For a standard thin viscous accretion disc around $M_z=10^5~\MSun$, $q=8$, $\chi_{1,2}=0.9$, and $t_c=4$ years MBHB, we can confidently measure (with a relative error of $<50 $ per cent) an Eddington ratio ${\rm f}_{\rm Edd}\sim0.1$ for a circular binary and ${\rm f}_{\rm Edd}\sim1$ for an eccentric system assuming $\mathcal{O}(10)$ stronger gas torque near-merger than at the currently explored much-wider binary separations. The minimum measurable eccentricity is $e_0\gtrsim10^{-2.75}$ in vacuum and $e_0\gtrsim10^{-2}$ in gas. A weak environmental perturbation (${\rm f}_{\rm Edd}\lesssim1$) to a circular binary can be mimicked by an orbital eccentricity during in-spiral, implying that an electromagnetic counterpart would be required to confirm the presence of an accretion disc.  
\end{abstract} 


\section{Introduction}

The prospect of the observation of gravitational waves (GWs) in the mHz band in the 2030s looks promising following the adoption by ESA of the Laser Interferometer Space Antenna (LISA; \citealt{AmaroSeoane2017,Barack2019,Colpi2024}) and with other projects, such as TianQin \citep{Wang2019} and Taiji \citep{Gong2021}, being developed. One of the primary expected extragalactic sources for LISA are massive black hole binaries (MBHBs) with primary-to-secondary mass ratios $q\lesssim10$ and total masses between $10^4~\MSun$ and $10^8~\MSun$, which LISA will be able to observe up to redshift $z\sim20$ \citep{AmaroSeoane2017}. Another expected source are intermediate/extreme mass ratio inspirals (I/EMRIs; \citealt{Babak2017,AmaroSeoane2018b}) with $q\gtrsim10^2$, which can be observed up to $z\lesssim2$. MBHBs, with their high signal-to-noise ratios (SNRs; \citealt{AmaroSeoane2017}), provide exciting opportunities to not only measure source properties with high accuracy but also place constraints on the properties of their environments.

The main formation channel for MBHBs is via galaxy mergers \citep{Begelman1980}. To shrink these binaries from a large scale to the coalescence phase within a Hubble time requires an environmental perturbation that could come from either gas or stars \citep[see, e.g.][]{AmaroSeoane2022}. In this paper, we will be primarily concerned with the dynamical effects of gas, as they can non-negligibly perturb both the semi-major axis and eccentricity of MBHBs in the LISA regime more strongly than stellar interactions, given the tight separations. Therefore, when we refer to an environment we will always mean a gas accretion disc. MBHs are often observed to be accompanied by an accretion disc at the center of active galactic nuclei (AGN) galaxies, especially beyond $z\gtrsim1$ and up to $z\lesssim7$ \citep{Padovani2017}. Therefore, as galaxy mergers trigger gas inflow and AGN activity \citep{Mayer2013}, MBHBs can be driven to coalescence by a surrounding gas reservoir. For the near-equal mass MBHBs considered here, we expect the accretion disc to take the form of a circumbinary disc (CBD; \citealt{DOrazio2015}).

GWs can be an important tool to not only measure the source properties but also probe imprints of the environment in which the binary is evolving. In the coalescence phase, gas mainly affects the binary via migration torques and mass accretion. The detectability of the imprint on the emitted waveform of these effects \citep[see, e.g.][]{Barausse2014,Garg2022,Caputo2020} is dependent on the details of the gas inflow. While most of the works on environmental measurements from GWs have focused on I/EMRIs \citep{Levin2007,Barausse2014,Derdzinski2019,Derdzinski2021,speri2023,Cole2023}, recent works have indicated that measurements of gas effects on more equal-mass MBHBs are also possible \citep{Garg2022,Dittmann2023,Tiede2023}. However, these studies for MBHBs have focused on the detectable accumulated dephasing in the GW waveform caused by gas, modelled using only Newtonian-order terms. This makes it optimistic as well as impossible to confidently pin down gas as a sole cause for this dephasing in the absence of an electromagnetic (EM) counterpart, since either higher-order post-Newtonian (PN) terms, eccentricity, or other environmental effects can also produce similar dephasings \citep[see, e.g.][]{Zwick2023}. Therefore, it becomes crucial to identify the region of parameter space in which we could confidently identify a gas accretion disc as the environment of a MBHB, utilizing only the observed emitted GWs.

While shrinking the MBHB from a large scale, gas can also excite eccentricity that can be measurable up to $\sim10^{-2.75}$ in the LISA band one year before the merger \citep{Garg2024}, despite partial circularization due to GWs \citep{Peters1963,Peters1964}. Depending on whether the CBD is prograde or retrograde and extremely or moderately thin, and whether the binary is equal-mass or unequal-mass, we can expect different eccentricities in the LISA band. Therefore, measurement of eccentricities can provide evidence towards certain disc configurations even if gas effects themselves become negligible near coalescence. However, if there are measurable gas effects, then there can be an interplay between them and the eccentricity when performing parameter estimation.

This work considers eccentric binaries of two aligned spinning MBHs embedded in a CBD. We aim to consider eccentricity as well as gas parameters during parameter estimation for either a one-year or four-year observation window. To be close to realistic data analysis methodologies, we use high-order post-Newtonian eccentric waveforms with aligned spin corrections to the circular part, we model LISA's motion and use the time delay interferometry response model, which will be needed to cancel the laser noise, and we consider both analytical and numerical techniques to assess the achievable constraints on the parameters of interest.

The paper is structured as follows. In Section~\ref{Sec:SPA}, we explain our methodology to include gas-induced corrections in the GW phase for eccentric MBHBs. Section~\ref{Sec:CBD} studies modeling of environmental effects from CBD simulations to get the leading-order dephasing from different gas effects. We summarize our parameters of interest and waveform model in Section~\ref{Sec:params}. In Section~\ref{Sec:Fisher}, we analytically compute errors on different parameters using the Fisher matrix formalism. We summarize our results from Bayesian inference in Section~\ref{Sec:MCMC}. In Section~\ref{Sec:Bias}, we study if a wrong template can mimic an injected signal. We discuss our findings in Section~\ref{Sec:discussion} and summarize the key takeaways of this work in Section~\ref{Sec:conclusion}.

\section{The stationary phase approximation}\label{Sec:SPA}

Let us consider two spinning BHs in vacuum at redshift $z$, with a redshifted total binary mass $M_z$ and a primary-to-secondary mass ratio $q\geq1$, revolving around each other in an eccentric orbit with their dimensionless spins\footnote{The dimensionless spin of a BH of mass $m$ and spin angular momentum $J$ is $\chi\equiv cJ/Gm^2$, where $G$ is the gravitational constant and $c$ is the speed of light.} $\chi_{1,2}$ aligned to the angular momentum of the binary. This is equivalent to the motion of a BH of reduced mass $\eta M_z$, where $\eta\equiv q/(1+q)^2$ is the symmetric mass ratio, in an elliptical orbit around a black hole of mass $M_z$, fixed at the focus.\footnote{See Tables~\ref{table:parameters} and \ref{table:Def_var} for definitions of commonly used variables and terms.} This orbit has a detector frame semimajor axis $a$, eccentricity $e$, and orbital angular frequency $\Omega$.

Due to the quadrupole nature of GW emission, the GW emission from a small eccentricity ($\lesssim 0.1$) and near-equal mass system is dominated by the $n=2$ eccentric harmonic, which is twice the orbital frequency:
\begin{equation}\label{eq:fa}
    f= \frac{1}{\pi}\left(\frac{GM_z}{a^3}\right)^\frac12,
\end{equation}
In the small-eccentricity limit, we can also approximate the orbital angular frequency as $\Omega=\pi f$ at all PN orders.

At the Newtonian-order,\footnote{Newtonian-order terms are denoted by the superscript (0).} the orbital averaged GW-driven semi-major axis decay rate is \citep{Peters1963,Peters1964}
\begin{align}\label{eq:aedot}
    \dot{a}^{(0)}_{\rm GW}&=-\frac{64}{5}\frac{G^3}{c^5}\frac{\eta M_z^3}{a^3}\text{F}(e),
\end{align}
where
\begin{align}
    \text{F}(e)&=\left(1+\frac{73}{24}e^2+\frac{37}{96}e^4\right)(1-e^2)^{-\frac72}.
\end{align}

The stationary phase approximation (SPA), which holds for slowly varying phase and amplitude over an orbital period \citep{Cutler1994}, is valid for the inspiral part of the GW signal. The SPA phase can be expressed as
\begin{align}\label{eq:SPA}
    \psi(f)&=2\pi f t_c-\phi_c+2\pi f\int\frac{da}{\dot{a}}-2\pi\int f\frac{da}{\dot{a}},
\end{align}
where $t_c$ and $\phi_c$ are the time and phase of coalescence, respectively.

For simplicity, we only consider circular Newtonian-order amplitude with the quadrupole mode (i.e., $(2,2)$ mode) for all cases. The phase is more sensitive than the amplitude to minor corrections arising from small eccentricity ($e\lesssim0.1$) \citep{Moore2016}, and the same should be true for weak environmental effects.

Since both eccentricity and environmental interactions affect the phase evolution of the source, we describe the cumulative phase of an event by its individual contributions: $\psi_{\rm TF2}$ is the phase a circular inspiral will accumulate in vacuum, solely due to emission of GWs; $\Delta\psi_{\rm TF2Ecc}$ represents the phase correction to an event's waveform when orbital eccentricity alters its GW emission; and $\Delta\psi_{\rm gas}$ represents the phase correction that is a consequence of environmental interaction, which further speeds up or slows down the inspiral. For the latter, we adopt the interaction with a gas disc as our fiducial environmental effect. As we discuss later on, $\Delta\psi_{\rm TF2Ecc}$ and $\Delta\psi_{\rm gas}$ have negligible cross-terms. We consider terms up to $3.5$PN order \citep{Buonanno2009} for $\psi_{\rm TF2}$ with aligned spin corrections also up to $3.5$PN order \citep{Arun2009,Mishra2016}, $3$PN and $\mathcal{O}(e^2)$ order for $\Delta\psi_{\rm TF2Ecc}$ \citep{Moore2016}, and the leading-order correction from gas in $\Delta\psi_{\rm gas}$. The total SPA phase can be expressed as
\begin{equation}
    \psi=2\pi f t_c-\phi_c+\psi_{\rm TF2}+\Delta\psi_{\rm TF2Ecc}+\Delta\psi_{\rm gas}.
\end{equation}

The overall semi-major axis evolution rate can be written down assuming no cross-term between GWs and gas effects\footnote{Thus far, most hydrodynamical simulations show that the gas torques and accretion rates are not strongly affected by GW-inspiral \citep{Tang2018,Derdzinski2019,Derdzinski2021}. However, there may be exceptions within the parameter space, which is yet to be fully explored.}:
\begin{align}\label{eq:adot}
\dot{a}&=\dot{a}_{\rm GW}+\dot{a}_{\rm gas}.
\end{align}

Given expected traditional disc model properties, we assume $\dot{a}_{\rm gas}\ll \dot{a}_{\rm GW}$ in the near-coalescence phase. Therefore, Eq.~\eqref{eq:SPA} can be expanded to separate the different contributions (all the integration constants are absorbed into $t_c$ and $\phi_c$):
\begin{align}\label{eq:PhaseBD}
    \psi_{\rm TF2}+ \Delta\psi_{\rm TF2Ecc} &= 2\pi f\int\frac{da}{\dot{a}_{\rm GW}}-2\pi \int f \frac{da}{\dot{a}_{\rm GW}},\\
    \Delta\psi_{\rm gas}&=-2\pi f\int da\frac{\dot{a}_{\rm gas}}{\dot{a}^2_{\rm GW}}+ 2\pi \int da~f\frac{\dot{a}_{\rm gas}}{\dot{a}^2_{\rm GW}}\label{eq:PhaseGas}.
\end{align}
The GR phasing contributions, $\psi_{\rm TF2}$ and $\Delta\psi_{\rm TF2Ecc}$ are well known in the literature. The gas contribution, $\Delta\psi_{\rm gas}$, appears here in its current form for the first time, although different forms of the same expressions do exist in the literature \citep[see, e.g.][]{Yunes2011}.

In the next section, we consider the evolution of an MBHB in the presence of an accretion disc to get an estimate of $\Delta\psi_{\rm gas}$.

\section{Modeling effects from a circumbinary disc}\label{Sec:CBD}
The torque exerted by a near-equal mass MBHB typically carves out a central cavity in the inner accretion disc that leads to the formation of a circumbinary disc (CBD; \citealt{DOrazio2015}). However, streams of gas still flow into this cavity, feeding mini-discs that accrete on to the binary as well as adding an additional torque component apart from an outer CBD \citep{Farris2014}. Non-axisymmetric features in this gas configuration exert a gravitational torque on the binary, which can lead to inspiral or outspiral of the MBHB depending on various disc-binary parameters. Typically a binary shrinks before GWs take over to drive it to merger, if we have a retrograde system \citep{Tiede2024}, or a prograde disc with an unequal mass binary \citep{Duffell2020}, a sufficiently thin disc \citep{Tiede2020}, or a moderately eccentric system \citep{DOrazio2021,Siwek2023}. In 2D CBD studies, this gas torque is usually expressed as a function of the accretion rate onto the binary\footnote{We can express $\dot{M}_z={\rm f}_{\rm Edd}M_z/\tau$, where ${\rm f}_{\rm Edd}$ is the Eddington ratio, and $\tau\approx50$Myr is the Salpeter timescale for our fiducial radiative efficiency of $0.1$.} ($\dot{M}_z$) and in the circular limit\footnote{All quantities in the circular limit have bar on the top.} as $\bar\Gamma_{\rm CBD}=\xi\dot{M}_za^2\Omega$, where the fudge factor $\xi$ depends upon the disc parameters and binary mass ratio \citep{Duffell2020,Garg2022}. This expression is consistent with the viscous torque estimate \citep{Lin1986}. Furthermore, $\xi$ tends to be positive (expand binary) for an equal-mass binary \citep{DOrazio2021} and negative (shrink binary) for $q\gtrsim10$ \citep{Cuadra2009,Moody2019,Munoz2019,Munoz2020,Duffell2020,Dittmann2023,Tiede2020,Tiede2024}. However, depending on the thermodynamic assumptions, we can also have a negative torque for a circular near-equal mass system \citep{Bourne2023}. Therefore, in this work, we will survey $\xi$ values that cover all realistic possibilities.

We consider the impact of both migration and mass accretion on the MBHB evolution in the following sections.

\subsection{Migration}
\label{sec:migration}
Studying the influence of gas-induced migration on the GW inspiral of the MBHB near merger has a lot of challenges. Gas effects on a tight near-equal mass circular MBHB have been simulated extensively, but only in the regime where GW inspiral is not important. Therefore, extrapolating results from these studies to near coalescence ($\dot{a}_{\rm gas}\ll \dot{a}_{\rm GW}$) could potentially lead to errors. A few studies on circular extreme-mass ratio BHBs embedded in a gas disc by \citet{Tang2018,Derdzinski2019,Derdzinski2021} find that gas effects do not change due to GW emission near the merger. However, they only consider Newtonian-order terms \citep{Peters1964} without including higher-order relativistic corrections. The inclusion of eccentricity could further exacerbate these problems. Hence, to encompass modeling uncertainties regarding gas migration effects on the embedded eccentric MBHB in the LISA band, we can write down a generic power law~\footnote{We found any cross terms between gas and eccentricity to be heavily suppressed in the phase, see Appendix~\ref{AppB}.}
\begin{align}\label{eq:adotgascir}
    \dot{\bar a}_\text{mig}=&{\mathcal{A}}\left(\frac{a}{GM_z/c^2}\right)^{n_{\rm g}}\dot{\bar a}^{(0)}_{\rm GW},
\end{align}
where dimensionless $\mathcal{A}$ and $n_{\rm g}$ are assumed to be constants for the duration of the binary coalescence time, $t_c$. Effectively, gas corrections to the SPA phase due to migration will enter at the $-n_{\rm g}$ PN order in this parametrization. Given that gas effects become increasingly negligible compared to GW emission towards the merger, we can safely assume $n_{\rm g}>0$. Moreover, we will only consider Newtonian-order hydrodynamical simulations\footnote{We note that relativistic corrections to the gas and binary motion can result in changes to the gas dynamics and resulting torque \citep{Berentzen2009,Wenshuai2021}, but here we focus on detecting gas with a more generic parameterization.} to study $\dot{a}_\text{mig}$. 

For $\dot{a}_{\rm gas}=\dot{\bar a}_{\rm mig}$ and assuming  $\dot{a}_{\rm GW}=\dot{a}^{(0)}_{\rm GW}$ in Eq.~\eqref{eq:PhaseGas}, we get the leading-order dephasing from migration \citep{Yunes2011}
\begin{align}\label{eq:phasemig3}
    \Delta\psi_{\rm mig}&=-\psi^{(0)}_{\rm TF2}\frac{{20\mathcal{A}}}{(n_{\rm g}+4)(2n_{\rm g}+5)}v^{-2n_{\rm g}}.
\end{align}

Assuming a thin CBD torque fiducial model, the torque on the binary in the circular limit is given by:
\begin{align}\label{eq:adotgascir2}
    \bar\Gamma_{\rm CBD}&=\xi\dot{M}a^2\Omega=\frac{d}{dt}(\eta M_z a^2\Omega),\nonumber\\ 
    \implies \dot{\bar a}_{\rm mig}&=\xi\frac{\dot{M}}{M_z}\frac{2}{\eta}a,
\end{align}
where in the first line we have ignored the mass accretion term when taking the time-derivative (see Section~\ref{subsec:massaccretion}). Therefore, equating Eqs~\eqref{eq:adotgascir} and \eqref{eq:adotgascir2} gives us
\begin{align}\label{eq:ampmig}
    {\mathcal{A}}&=-5.40\times10^{-15}\frac{\xi}{1.0}\frac{{\rm f}_{\rm Edd}}{1.0}\frac{0.1}{\epsilon}\left(\frac{\eta}{0.1}\right)^{-2}\frac{M_z}{10^5\MSun},\\
    n_{\rm g}&=4\nonumber.
\end{align}

Plugging $\mathcal{A}$ and $n_{\rm g}$ into Eq.~\eqref{eq:phasemig3} gives us
\begin{align}\label{eq:phasemig}
     \Delta\psi_{\rm mig}&=\psi^{(0)}_{\rm TF2}C_{\rm mig}v^{-8},\\
     C_{\rm mig}&=1.04\times10^{-15}\frac{\xi}{1.0}\frac{{\rm f}_{\rm Edd}}{1.0}\frac{0.1}{\epsilon}\left(\frac{\eta}{0.1}\right)^{-2}\frac{M_z}{10^5\MSun}\nonumber,
\end{align}
where $v\equiv(GM_z\pi f/c^3)^{\frac13}$ is the characteristic velocity of the binary, $\psi^{(0)}_{\rm TF2}\equiv(3/128\eta)v^{-5}$ is the leading-order phase term.

\subsection{Mass accretion}\label{subsec:massaccretion}

The increase in mass of either BH due to mass accretion can also affect the binary evolution. Heavier BHs will coalesce faster due to stronger GW emission as per Eq.~\eqref{eq:aedot}. Preferential accretion onto one of the BHs will change the center of mass equilibrium, and accretion of gas with linear or angular momentum will change a BH's momentum and spin, respectively. Each of these effects produces an additional effective torque, which can be comparable to or much weaker than the gravitational component discussed in Section \ref{sec:migration}. We neglect these components in this study, given that we only consider accretion onto the total binary and not onto its individual components, nor its small-scale gas configuration. We note however that the inclusion of these effects may justify an increase in the torque amplitude $C_{\rm mig}$ in Eq.~\eqref{eq:phasemig}.

The accretion rate onto the binary for $\epsilon=0.1$ can be expressed as
\begin{align}\label{eq:massacc}
    \dot{M}_z=&{\rm f}_{\rm Edd}M_z/\tau\nonumber,\\
    \implies M_z=&M_{z,0}\exp({\rm f}_{\rm Edd}t/\tau).
\end{align}

The phase contribution due to mass accretion in the circular limit can be expressed as \citep{Caputo2020}:
\begin{align}\label{eq:phaseacc2}
    \Delta\psi_{\rm acc}=\psi^{(0)}_{\rm TF2}\frac{25}{256}\frac{{\rm f}_{\rm Edd}}{\tau}\frac{GM_z\eta^{-1}}{c^3}\left(\frac{1}{3}v_0^{-8}-\frac{15}{26}v^{-8}\right),
\end{align}
where we have absorbed terms independent of $f$ into $\phi_c$ and terms proportional to $f$ in $t_c$. We have replaced $M_{z,0}$ with $M_z$ after computations, since the LISA observation duration we consider is at maximum $4~{\rm years}$, implying ${\rm f}_{\rm Edd}t/\tau\ll1$ in Eq.~\eqref{eq:massacc}. First term in $\Delta\psi_{\rm acc}$ can be re-expressed as
\begin{align}
    \Delta\psi^{(I)}_{\rm acc}&\approx\psi^{(0)}_{\rm TF2}
  \times10^{-16}\frac{{\rm f}_{\rm Edd}}{\eta}\frac{M_z}{10^5\MSun}\left(\frac{a_0}{GM/c^2}\right)^4\nonumber,
\end{align}
where $a_0$ is the initial separation, which cannot be more than $\sim\mathcal{O}(100GM_z/c^2)$ to have the MBHB merge within the LISA observation window. Therefore, $\Delta\psi^{(I)}_{\rm acc}$ is at maximum $\mathcal{O}(10^{-6})\psi^{(0)}_{\rm TF2}$ even for optimistic accretion rates (${\rm f}_{\rm Edd}$ $\lesssim100$) for near equal-mass MBHBs. Hence, the first term in Eq.~\eqref{eq:phaseacc2} can be dropped due to being negligible with respect to the leading-order SPA phase contribution $\psi^{(0)}_{\rm TF2}$ and we get the effective accretion dephasing
\begin{align}\label{eq:phaseacc}
    \Delta\psi_{\rm acc}&=\psi^{(0)}_{\rm TF2}C_{\rm acc}v^{-8},\\
    C_{\rm acc}&=-1.96\times10^{-16}\frac{{\rm f}_{\rm Edd}}{1.0}\frac{0.1}{\epsilon}\left(\frac{\eta} {0.1}\right)^{-1}\frac{M_z}{10^5\MSun}.\nonumber
\end{align}

\subsection{Effective dephasing due to gas}
Both phasing terms from migration and accretion have the same $v^{-8}$ frequency dependence with respect to the vacuum GW phase, which implies that gas corrections to the SPA phase enter at the $-4$PN order. The amplitudes ($C_{\rm mig}$ and $C_{\rm acc}$) of both effects have similar dependencies on the binary-disc parameters. However, $C_{\rm mig}/C_{\rm acc}\sim(0.25/\eta)$ implies that dephasing due to migration is only comparable to accretion dephasing for equal-mass and becomes increasingly weaker for higher mass ratio binaries.

The overall phasing contribution from gas from both migration and accretion can be expressed as
\begin{equation}\label{eq:phasegas2}
    \Delta\psi_{\rm gas}=C_{\rm g}\psi^{(0)}_{\rm TF2}v^{-2n_{\rm g}}\left(\frac{\eta}{0.1}\right)^{-2}\frac{M_z}{10^5\MSun},
\end{equation}
where $C_{\rm g}$ and $n_{\rm g}$ depend on the underlying gas model. For our CBD torque fiducial model, we have:
\begin{align}\label{eq:Cg}
    C_{\rm g}&=10^{-15}\frac{\xi}{1.0}\frac{{\rm f}_{\rm Edd}}{1.0}\frac{0.1}{\epsilon},\\
    n_{\rm g}&=4.
\end{align}
We model $\Delta\psi_{\rm gas}$ based on the migration dephasing in Eq.~\eqref{eq:phasemig} and any uncertainties about the dependence on the binary-disc parameters are folded into $\xi$, which can either be positive or negative.

\section{Parameter space, LISA response, and time delay interferometry}\label{Sec:params}
To study the evolution of a MBHB embedded in an accretion disc, we mainly consider the binary-disc parameters summarized in Table~\ref{table:parameters}, which are defined in the detector frame. There are eight intrinsic parameters (first eight rows of Table~\ref{table:parameters}) and six extrinsic parameters (last six rows of Table~\ref{table:parameters}). We further divide the intrinsic parameters into five intrinsic-merger $\{M_z,q,\chi_1,\chi_2,t_c\}$ and three intrinsic-inspiral $\{e_0,C_{\rm g},n_{\rm g}\}$ parameters, due to their relative importance in different phases of the GW source evolution.

\begin{table}
\centering
    \begin{tabular}{|C|p{0.5\linewidth}|c|}
        \hline
        \pmb{\theta}&{\bf Definition} & {\bf Units}\\
        \hline
        \hline
        M_z &Total redshifted mass & $\MSun$\\
        \hline
        q & Mass ratio& Dimensionless\\
        \hline
        \chi_{1,2}&Spin parameters of both BHs & Dimensionless\\
        \hline
         t_c &Time of coalescence& year\\
        \hline
        e_0 & Initial eccentricity & Dimensionless\\
        \hline
        C_{\rm g}&Environmental amplitude  & Dimensionless\\
        \hline
        n_{\rm g} & Environmental semi-major axis power-law relative to GWs & Dimensionless\\
        \hline
        D_{\rm L}&Luminosity distance & Mpc\\
        \hline
        \phi_c&Phase at coalescence & Radian\\ 
        \hline
        \imath & Inclination & Radian \\ 
        \hline
        \lambda  & Ecliptic latitude & Radian \\ 
        \hline
         \beta&Ecliptic longitude & Radian \\ 
        \hline
         \psi&Initial polarization angle & Radian \\ 
        \hline
    \end{tabular}
\caption{Parameters of interest in the detector frame.}
\label{table:parameters}
\end{table}

We add the gas-induced phasing term in Eq.~\eqref{eq:phasegas2} to the \textsc{TaylorF2Ecc} \citep{Moore2016} phase and then modify \textsc{lisabeta} \citep{Marsat2021} to incorporate this waveform model. \textsc{lisabeta} takes into account LISA's motion and computes the time delay interferometry (TDI) response of the detector. We use the LISA sensitivity curve including galactic confusion noise from \citet{Marsat2021}.

Based on realistic astrophysical expectations for the different parameters in Table~\ref{table:parameters}, we generate waveforms until the innermost stable circular orbit over these parameter grids:
\begin{align}\label{eq:parameters}
    M_z\in&\{10^{4.5},10^{5},10^{5.5},10^{6}\}~\MSun,\nonumber\\
    q\in&\{8,1.2\}\iff\eta\in\{0.1,0.25\},\nonumber\\
    \chi_{1,2}\in&\{0.9\},\nonumber\\
    e_0\in&\{0,10^{-3},10^{-2.75},10^{-2.5},10^{-2.25},10^{-2},10^{-1.75},\nonumber\\
    &10^{-1.5},10^{-1.25},0.1\}\nonumber,\\
    t_c\in&\{1~{\rm year}, 4~{\rm years}\},\nonumber\\
    C_{\rm g}[10^{-15}]\in&\{-10^3,-10^2,-10^1,-10^0,-0.1,0,0.1,10^0,10^1,\nonumber\\
    &10^2,10^3\}\nonumber,\\
    n_{\rm g}\in&\{4\}.
\end{align}
The extrinsic fiducial parameters are $z=1$, which corresponds to $D_{\rm L}=6791.3$ Mpc for the best-fit \citet{Planck2020} cosmology, and all angles are set to $0.5$ radians. The systems we consider here spend at least 4 years in the LISA band before merging. We limit $e_0\geq10^{-3}$, since eccentricities below that are unmeasurable for MBHBs \citep{Garg2024}. Considering that BHs in gas are expected to be highly spinning \citep[see, e.g.][]{Reynolds2021}, we chose $\chi_{1,2}=0.9$ as our fiducial case. However, we find that our constraints on intrinsic-inspiral parameters are almost independent of the exact spin magnitude. The SNRs of the sources in the grid specified in~\eqref{eq:parameters} range from $\sim150$ to $\sim2500$ at $z=1$ with tiny differences for the two times of coalescence.\footnote{See Fig.~2 of \citet{Garg2024} for SNRs measured in a one year observation for our systems of interest at different redshifts.} 
In the next section, we compute relative errors on the measurement of parameters.

\section{Fisher formalism}\label{Sec:Fisher}
\subsection{Fisher matrix measurement of parameters for a $M_{\lowercase{z}}=10^5~\MSun$ and $\lowercase{q}=8.0$ MBHB with $\lowercase{t}_{\lowercase{c}}=4$ years}\label{Sec:Fisher1}
Using the Fisher formalism \citep{Vallisneri2008}, we compute the expected error ($\sigma^{\rm Fisher}_\theta$) on each parameter $\theta$ and set a threshold on the relative error $\delta\theta[\%]\equiv100*\sigma^{\rm Fisher}_\theta/|\theta|<50$ to define when a parameter is considered to be measured. We use $50$ per cent as a threshold since we are interested in determining whether the parameter differs from zero, i.e., whether an effect is present, rather than wanting an accurate measurement of a large effect. We can use the Savage-Dickey ratio \citep{Dickey1971} to convert our Fisher-based error threshold to the more customary Bayes factor ($\mathcal{B}$). For a  recovered value $k\sigma$ away from zero assuming a  Gaussian posterior with standard deviation $\sigma$ and a uniform prior of width $l\sigma$ on the parameter of interest, implies a Bayes factor of \citep{Dickey1971,Garg2024}
\begin{equation}
    \mathcal{B}=\frac{1}{l\sigma}\sqrt{2\pi\sigma^2}\exp\left({\frac12\frac{(k\sigma)^2}{\sigma^2}}\right)=\frac{\sqrt{2\pi}}{l}\exp\left(\frac{k^2}{2}\right).
\end{equation}
For a $50$ per cent threshold, i.e., $k=2$ and 
assuming $l=5$, which ensures the prior is just big enough to contain the entire region of high likelihood support and hence is the largest Bayes factor that would not be prior-dominated, gives $\ln\mathcal{B}\approx1.3$, which has a substantial strength \citep{Taylor2021}. To get decisive strength ($\ln\mathcal{B}\gtrsim5$), we would need a relative error below $30$ percent or $k\gtrsim3.3$.

We first consider a MBHB with $M_z=10^5~\MSun$ and $q=8.0$ and with a coalescence time of $t_c=4$ years, which has an SNR of $377.85$ and generates $19159$ GW cycles in the LISA band. We study both circular and eccentric (with $e_0=0.01$ or $e_0=0.1$) cases. In this section, we always keep the intrinsic-merger parameters $\{M_z,q,\chi_1,\chi_2,t_c\}$ free and we show relative errors in vacuum in Table~\ref{table:Fisher_Err_Int}. Unsurprisingly, including eccentricity as a free parameter increases the uncertainties on all the intrinsic-merger parameters. Furthermore, given that we expect our fiducial CBD system to have the environmental power-law $n_{\rm g}=4$ and $n_{\rm g}$ has a non-Gaussian posterior, based on Fig.~\ref{fig:MCMC_4yr_ng} discussed in Section~\ref{Sec:MCMC}, we always keep it fixed to its fiducial value. 

\begin{table}
\centering
    \begin{tabular}{|C|C|C|C|}
        \hline
        \pmb{\delta\theta[\%]} & \pmb{\rm Circular}&\pmb{e_0=0.01}&\pmb{e_0=0.1}\\
        \hline
        \hline
        \delta M_z[\%]/10^{-2} & 3.41& 3.97&3.97\\
        \hline
        \delta q[\%]/10^{-2} &7.30 & 8.51&8.52\\
        \hline
        \delta \chi_1[\%]/10^{-2} & 6.52& 6.58&6.56\\
        \hline
        \delta \chi_2[\%]/10^{-1} & 9.04& 9.56&9.51\\
        \hline
         \delta t_c[\%]/10^{-7}& 2.67&  3.08&3.05\\
         \hline
         \delta e_0[\%]/10^{0}& - &  1.41&0.01\\
        \hline
    \end{tabular}
\caption{The relative uncertainties, computed using the Fisher formalism,  on the intrinsic-merger parameters and eccentricity for a vacuum system with $M_z=10^5~\MSun$, $q=8.0$, $\chi_{1,2}=0.9$, and $t_c=4$ years between three eccentricities: $e_0=0,~e_0=0.01$ and $e_0=0.1$.}
\label{table:Fisher_Err_Int}
\end{table}

\subsubsection{Circular MBHB}

\begin{figure}
    \centering
    \includegraphics[width=0.5\textwidth]{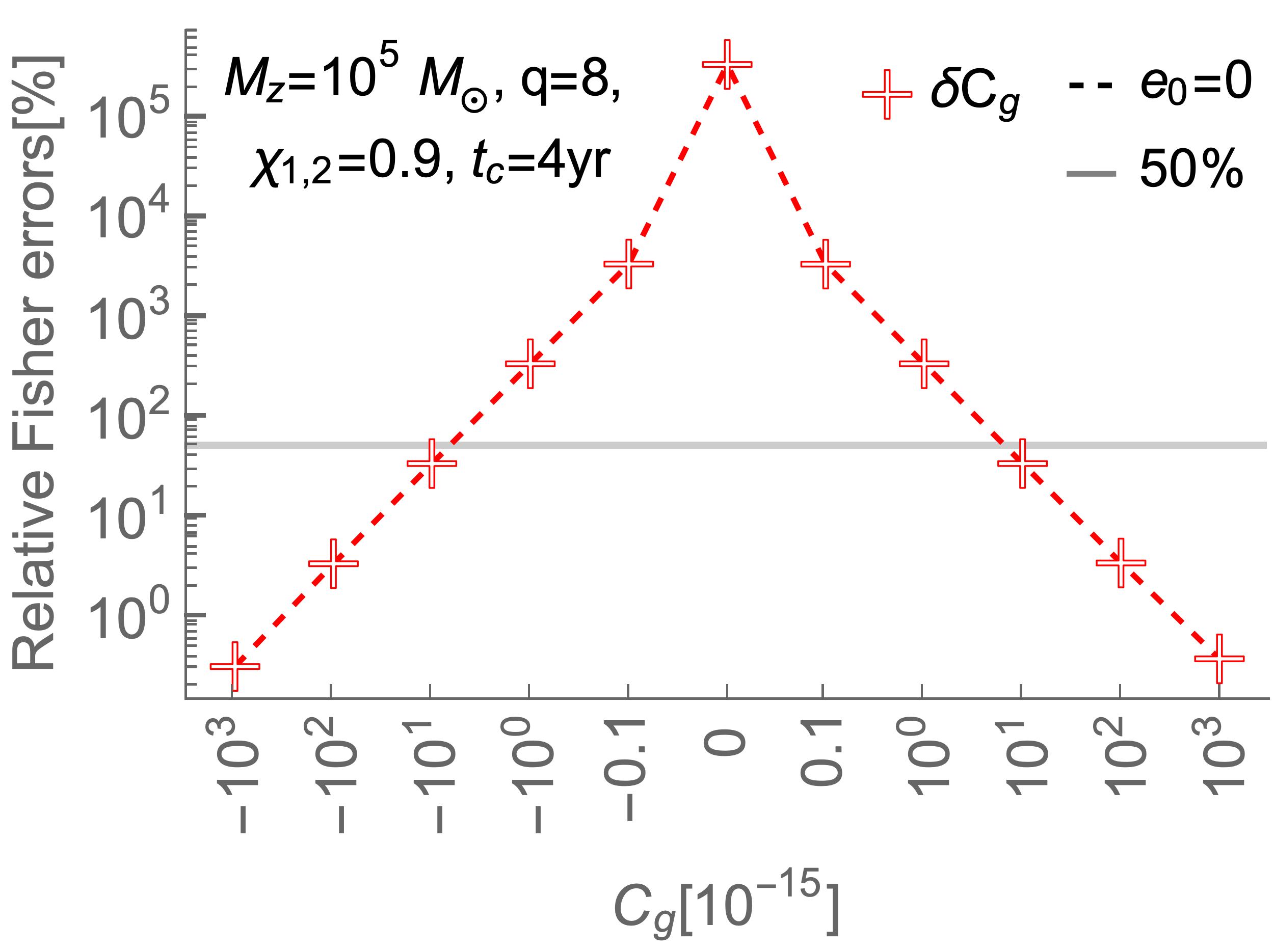}
    \caption{The relative error on the measurement of the environmental effective amplitude $\delta C_{\rm g}[\%]$ as a function of $C_{\rm g}$, denoted by a red `+' symbol, assuming zero eccentricity and fixed environmental power-law $n_{\rm g}$. The dashed line represents the circular case. The $C_{\rm g}=0$ case is computed using $C_{\rm g}=10^{-18}$ for numerical reasons. We denote the $50$ per cent well-measured threshold by a solid gray line.}
    \label{fig:Envamp_ecc0}
\end{figure}

We compute relative errors in the presence of an environment for a circular system. In Fig.~\ref{fig:Envamp_ecc0}, we show results for $\delta C_{\rm g}[\%]$ as a function of $C_{\rm g}$. We find that $\delta C_{\rm g}[\%]\approx5.34\times10^2/|10^{15}C_{\rm g}|$, almost independently of the sign of the environmental amplitude. The absolute error $\sigma^{\rm Fisher}_{C_{\rm g}}$ is independent of the value of $C_{\rm g}$, as expected due to the phase correction from the gas being linearly dependent on $C_{\rm g}$. Moreover, we can well-measure $|C_{\rm g}|\gtrsim 10^{-14}$.

The relative errors on the intrinsic-merger parameters with the inclusion of gas are higher than the values in vacuum given in Table~\ref{table:Fisher_Err_Int}, but are independent of the magnitude of $C_{\rm g}$. Therefore, the change in uncertainties is due to having an extra free parameter in the model. 

\subsubsection{Eccentric MBHB}
\begin{figure}
    \centering
    \includegraphics[width=0.5\textwidth]{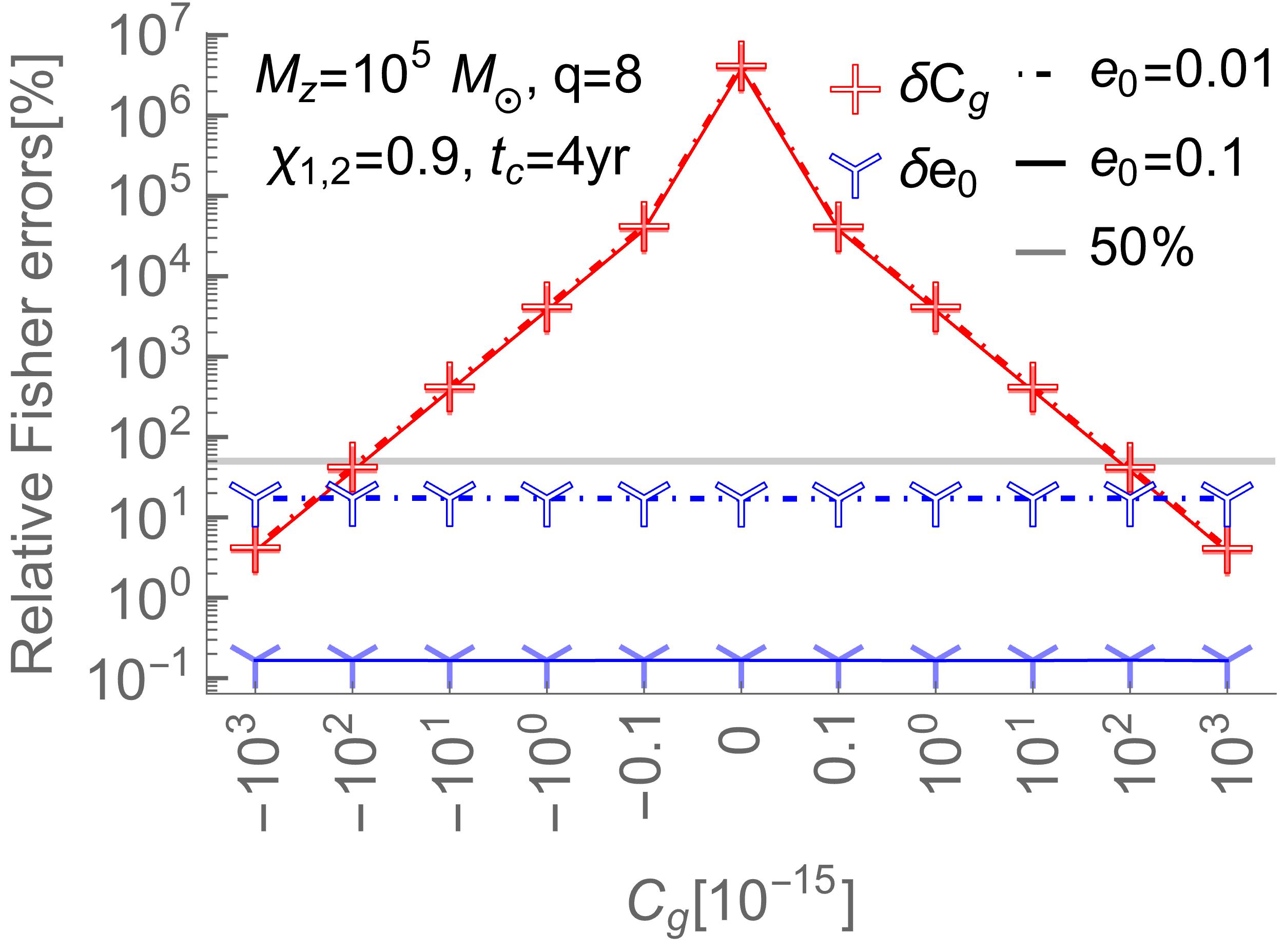}
    \caption{Same as in Fig.~\ref{fig:Envamp_ecc0}, but now including relative errors $\delta e_0$ for systems with initial eccentricities $e_0=0.01$ and $e_0=0.1$, denoted by a blue `$\Ydown$' symbol. We indicate the $e_0=0.01$ and $e_0=0.1$ cases by dot-dashed and solid lines, respectively. By including eccentricity, the relative errors on $C_{\rm g}$ increases by almost an order of magnitude with respect to Fig.~\ref{fig:Envamp_ecc0}.}
    \label{fig:Envamp_ecc}
\end{figure}

In the presence of an environment, we compute relative uncertainties for injected signals with eccentricities $e_0=0.01$ and $e_0=0.1$. In Fig.~\ref{fig:Envamp_ecc}, we show results for $\delta C_{\rm g}[\%]$ and $\delta e_0[\%]$ as a function of $C_{\rm g}$. We find that $\delta C_{\rm g}[\%]\approx6.35\times10^3/|10^{15}C_{\rm g}|$, irrespective of the eccentricity value, as expected, but higher than the circular case shown in Fig.~\ref{fig:Envamp_ecc0}. The eccentricity uncertainty, $\delta e_0[\%]$, is nearly independent of the environmental amplitude, but reaches higher values than for the vacuum case shown in Table.~\ref{table:Fisher_Err_Int}. Both increases can be attributed to having an extra free intrinsic-inspiral parameter. $|C_{\rm g}|\gtrsim10^{-13}$ is well-measured for our fiducial system.

The relative errors on intrinsic-merger parameters are again almost independent of the magnitude of $C_{\rm g}$ but are higher than the values for eccentric vacuum systems shown in Table~\ref{table:Fisher_Err_Int}, because of the extra free parameter in the model.

\subsubsection{Summary}
The major conclusions that can be drawn from this Section are:
\begin{itemize}
    \item All relative errors are nearly independent of the sign of the environmental amplitude $C_{\rm g}$.
    \item The relative errors on the intrinsic-merger parameters, $\{M_z,q,\chi_1,\chi_2,t_c\}$, and eccentricity, $e_0$, are independent of the value of $C_{\rm g}$, i.e., they remain almost the same for any value of $C_{\rm g}$. Moreover, they only increase slightly with respect to their values in vacuum when the environmental effect is included in the model. This is mainly because of the inclusion of an extra free parameter.
    \item The relative errors on $C_{\rm g}$ are higher in the eccentric case than in the circular case. However, the change is the same for both $e_0=0.01$ and $e_0=0.1$. Again, this is because of the inclusion of an extra degree of freedom in the eccentric environmental model with respect to the circular environmental one.
    \item For all the scenarios explored here, amplitudes less than $C_{\rm g}]\lesssim10^{-14}$ are not well measured.
\end{itemize}
In the next section, we will extend the Fisher results to cover the full parameter space defined in~\eqref{eq:parameters}. Based on the results in this section, we will only consider $C_{\rm g}\geq10^{-15}$ and choose $C_{\rm g}=10^{-13}$ and $e_0=0.1$ as our fiducial intrinsic-inspiral parameters.

\subsection{Fisher matrix measurement of parameters for systems of interest}\label{Sec:Fisher2}
We would now like to explore the parameter space of Eq.~\eqref{eq:parameters}. For this, we will make matrix plots for the environmental amplitude $C_{\rm g}$ and the initial orbital eccentricity $e_0$ by always considering a fixed environmental power-law $n_{\rm g}=4$. We are again keeping the intrinsic-merger parameters $\{M_z,q,\chi_1,\chi_2,t_c\}$ free.

\subsubsection{$\delta C_{\rm g}[\%]$}
\begin{figure}
    \centering
    \includegraphics[width=0.5\textwidth]{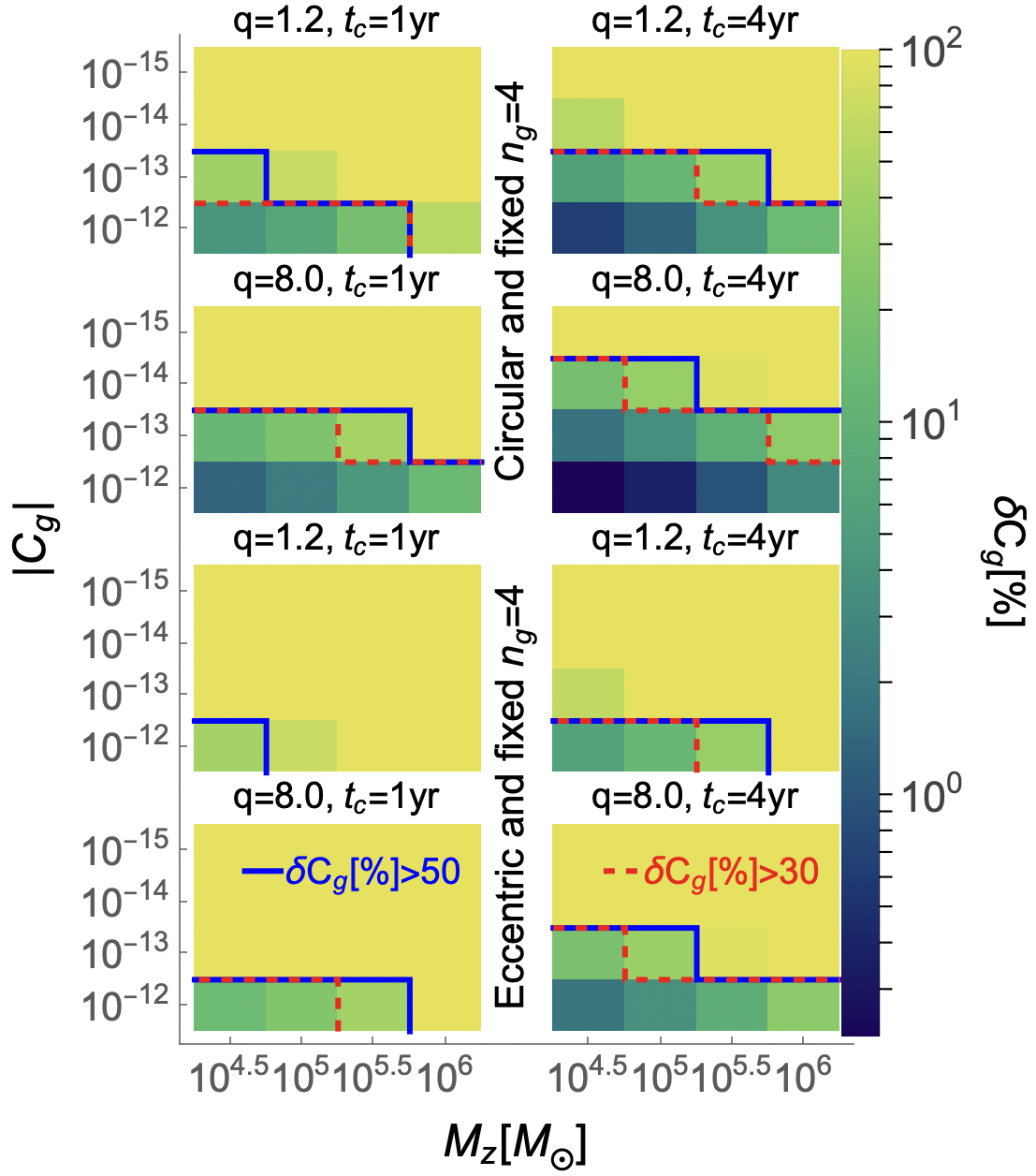}
    \caption{The relative Fisher errors on the environmental amplitude, $\delta C_{\rm g}[\%]$, as a function of $|C_{\rm g}|$ and the total mass $M_z$. The environmental power-law $n_{\rm g}$ is fixed to $4$ in all panels. In all eight panels, we vary $M_z$ from $10^{4.5}$ to $10^6~\MSun$ and $C_{\rm g}$ from $10^{-12}$ to $10^{-15}$. For all the left panels, we set $t_c=1$ year and assume $t_c=4$ years for all the right panels. We have either $q=1.2$ (in the first and third rows) or $q=8.0$ (in the second and fourth rows). In the top four panels, we assume circular orbits and in the bottom four panels we set $e_0=0.1$, and allow the initial eccentricity $e_0$ to also be a free parameter. We draw a red dashed line and a solid blue line to identify the region of well-measured parameters which have relative errors below $30$ and $50$ per cent, respectively. Moreover, we suppress all errors beyond $100$ per cent. We can at best well measure $C_{\rm g}\gtrsim10^{-14}$ for circular binaries with $q=8$ and $t_c=4$ years and are only able to constrain $C_{\rm g}=10^{-12}$ for an eccentric system with $q=1.2$ and $t_c=1$ year.}
    \label{fig:Envamp_Matrixplot}
\end{figure}

We want to study the uncertainty in the environmental amplitude $C_{\rm g}$ as a function of $C_{\rm g}$ for both circular and eccentric systems. Based on the findings of Section~\ref{Sec:Fisher1}, we only need to consider positive values of $C_{\rm g}\in\{10^{-12},10^{-13},10^{-14},10^{-15}\}$ and choose one eccentricity (here $e_0=0.1$). In Fig.~\ref{fig:Envamp_Matrixplot}, we show $\delta C_{\rm g}[\%]$ for various different choices of the intrinsic-merger parameters and strength of the environmental effect. Since $M_z$ defines if the inspiral part of the signal is in the low or high sensitivity region of the LISA frequency band, it affects the uncertainty on $C_{\rm g}$. Moreover, $q$ and $t_c$ set the number of GW cycle and so they also affect the measurement of the environmental amplitude. Based on the parameter dependence in Eq.~\eqref{eq:Cg}, for $M_z=10^6~\MSun$, $q=8.0$ or $\eta=0.1$, and $\xi=100$, we can confidently measure ${\rm f}_{\rm Edd}\gtrsim0.1$ for circular systems and ${\rm f}_{\rm Edd}\gtrsim1$ for eccentric MBHBs. These constraints on ${\rm f}_{\rm Edd}$ depend upon setting $\xi=100$, assuming that gas torques become $\mathcal{O}(10)$ stronger near-merger than at the currently explored wider separations where GWs are not important and extrapolating of the simulation results of \citet{Dittmann2022}, who found increasingly strong torques on binaries for thinner discs (see their Fig.~3).

\subsubsection{$\delta e_0[\%]$}
\begin{figure}
    \centering
    {\includegraphics[width=0.5\textwidth]{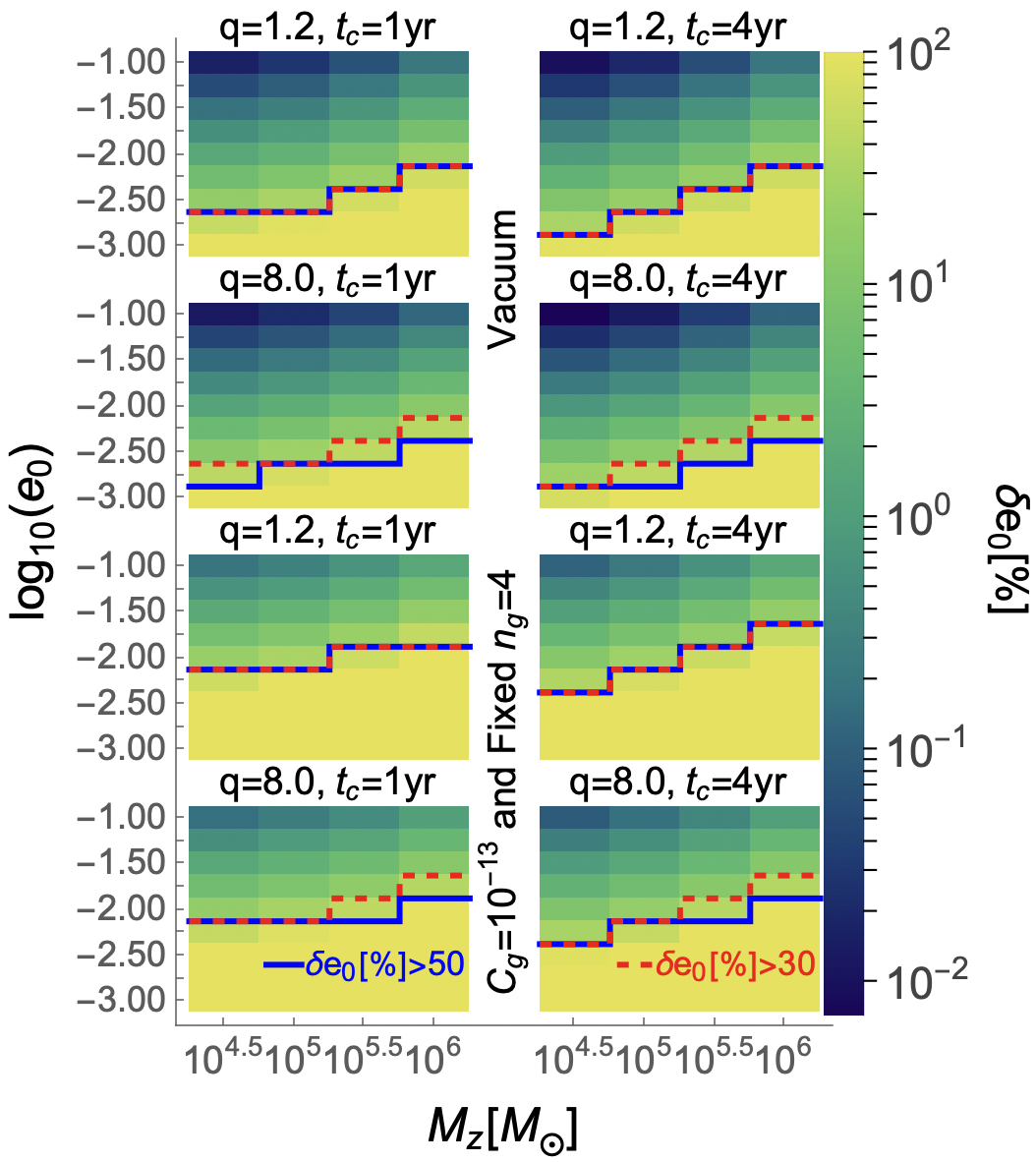}}
    \caption{The relative error on the initial eccentricity, $\delta e_0[\%]$, as a function of $e_0$ and $M_z$. In all eight panels we vary eccentricities from $10^{-3}$ to $0.1$ and masses from $10^{4.5}$ to $10^6~\MSun$. In the top four panels, we assume vacuum and in the bottom four panels we have $C_{\rm g}=10^{-13}$ and fixed $n_{\rm g}=4$. All other parameter choices are the same as in Fig.~\ref{fig:Envamp_Matrixplot}. In vacuum, we are able to measure at minimum $e_0\gtrsim10^{-2.75}$. While in the presence of an environment, at best we can measure $e_0\gtrsim10^{-2}$.}
    \label{fig:Ecc_Matrixplot}
\end{figure}

Now we want to compute the relative errors on the initial eccentricity, $e_0$, as a function of $e_0$, with or without the environmental effect. In Fig.~\ref{fig:Ecc_Matrixplot}, we show $\delta e_0[\%]$ in vacuum and in the presence of an environment with $C_{\rm g}=10^{-13}$. The measurement on $e_0$ strongly depends upon $M_z$ for a similar reason to the dependence of $\delta C_{\rm g}[\%]$ upon $M_z$. It also varies weakly with $q$ and $t_c$, since limited information is contained in the GW cycles observed from the very early inspiral.
For $t_c=1$ year we find a higher minimum measurable eccentricity in vacuum than what is reported in \citet{Garg2024} because of the extra number of free intrinsic-merger parameters $\{\chi_{1},\chi_{2},t_c\}$ in the analysis.

In the next section, we will carry out Bayesian inference to verify the results of this section and demonstrate the validity of our Fisher analysis.

\section{Bayesian inference}\label{Sec:MCMC}
We use the exact same Bayesian setup outlined in Section~4 of \citet{Garg2024} to find posterior distributions for a few systems of interest. Salient features include a zero noise realization and the Fisher initialization\footnote{starting walkers inside a multivariate Gaussian with its mean given by the injected values and covariance by the Fisher matrix.} to accelerate the parallel tempering Markov-chain Monte Carlo (PTMCMC) runs using the $\textsc{ptmcmc}$\footnote{https://github.com/JohnGBaker/ptmcmc} sampler. We set uniform priors for all parameters. The resulting posteriors can be used to cross-verify the Fisher matrix results reported in Section~\ref{Sec:Fisher}. 

\begin{figure*}
    \centering
    \includegraphics[width=\textwidth]{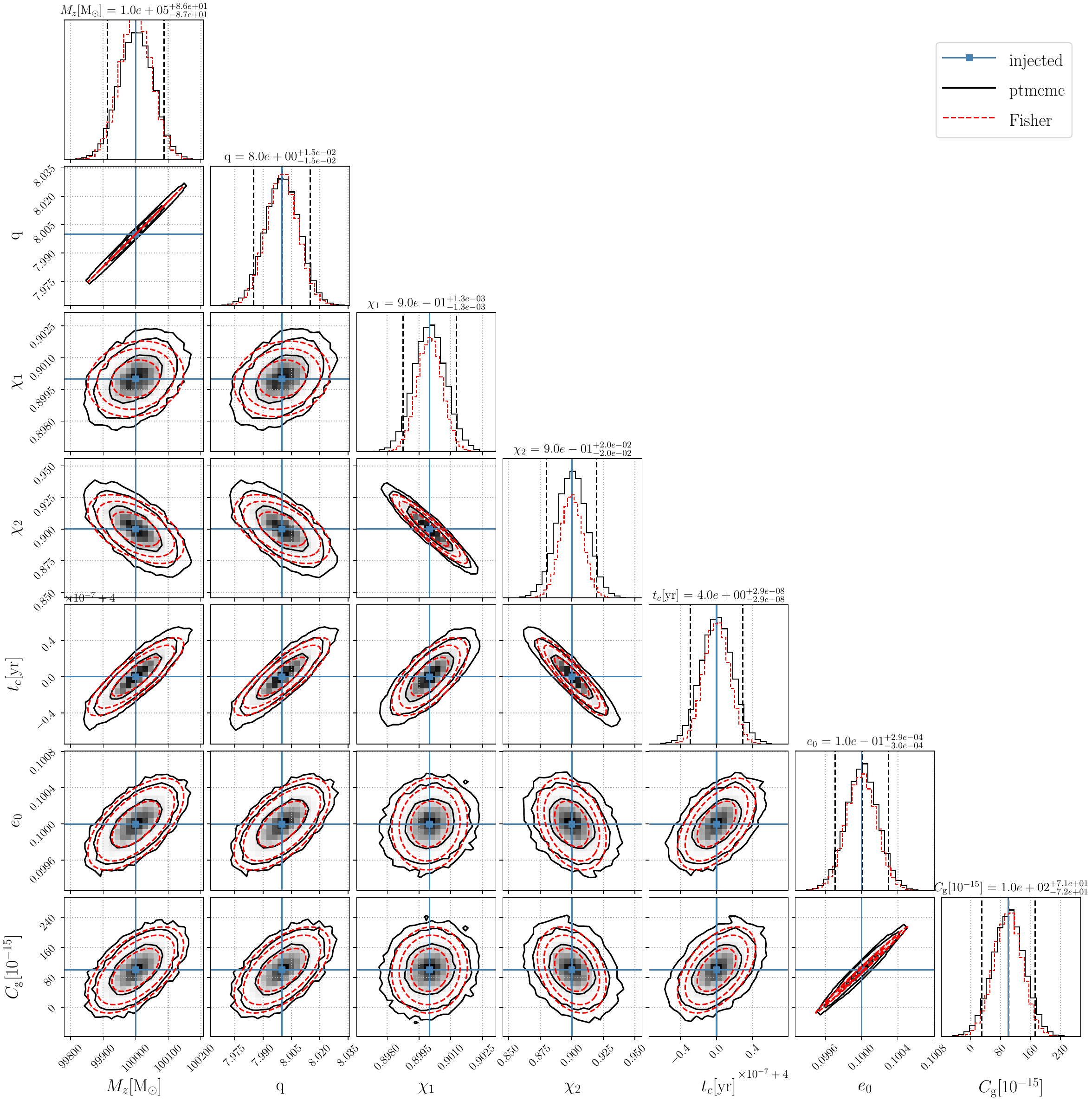}
    \caption{Posterior distributions (solid black) for a gas-embedded eccentric MBHB with parameters $M_z=10^5~\MSun$, $q=8.0$, $\chi_{1,2}=0.9$, and $t_c=4$ years with an effective gas-amplitude $C_{\rm g}=10^{-13}$ and a fixed environmental power-law $n_{\rm g}=4$. The blue lines mark the injected values, the middle dashed line shows the median of the distribution, and the two extreme vertical dashed lines indicate the $90$ per cent symmetric credible interval. 
    The two-dimensional contours of the posteriors indicate $68$, $95$, and $99$ per cent credible intervals. We also show the Fisher matrix predictions (dashed red) for comparison.}
    \label{fig:MCMC}
\end{figure*}

First, we show posteriors for our fiducial system in Fig.~\ref{fig:MCMC}.\footnote{The posterior for $t_c=1$ year is similarly Gaussian as in Fig.~\ref{fig:MCMC}, although with higher covariances as expected.} We can infer that all the parameters are well-recovered and the posteriors almost overlap with the Fisher matrix predictions. We conclude that our Fisher results in Section~\ref{Sec:Fisher} are robust. \footnote{We also show how posteriors on $C_{\rm g}$ change for different eccentricities in Fig.~\ref{fig:Comp_Cg100}.} Moreover, the degeneracy between $e_0$ and $C_{\rm g}$ is due to both being inspiral-only effects that decay as the separation decreases. The eccentricity decreases because of the GW-induced circularization and environmental perturbation decays due to entering phase at the $-4$PN order. Over the time that these effects are significant, the orbital velocity does not evolve very much, which allows effects with different post-Newtonian orders to compensate for each other. This means that very similar waveforms are produced for different combinations of $e_0$ and $C_{\rm g}$. For instance, increasing environmental amplitude to a higher positive value can be offset by having a higher eccentricity. 

\begin{figure}
    \centering
    \includegraphics[width=0.5\textwidth]{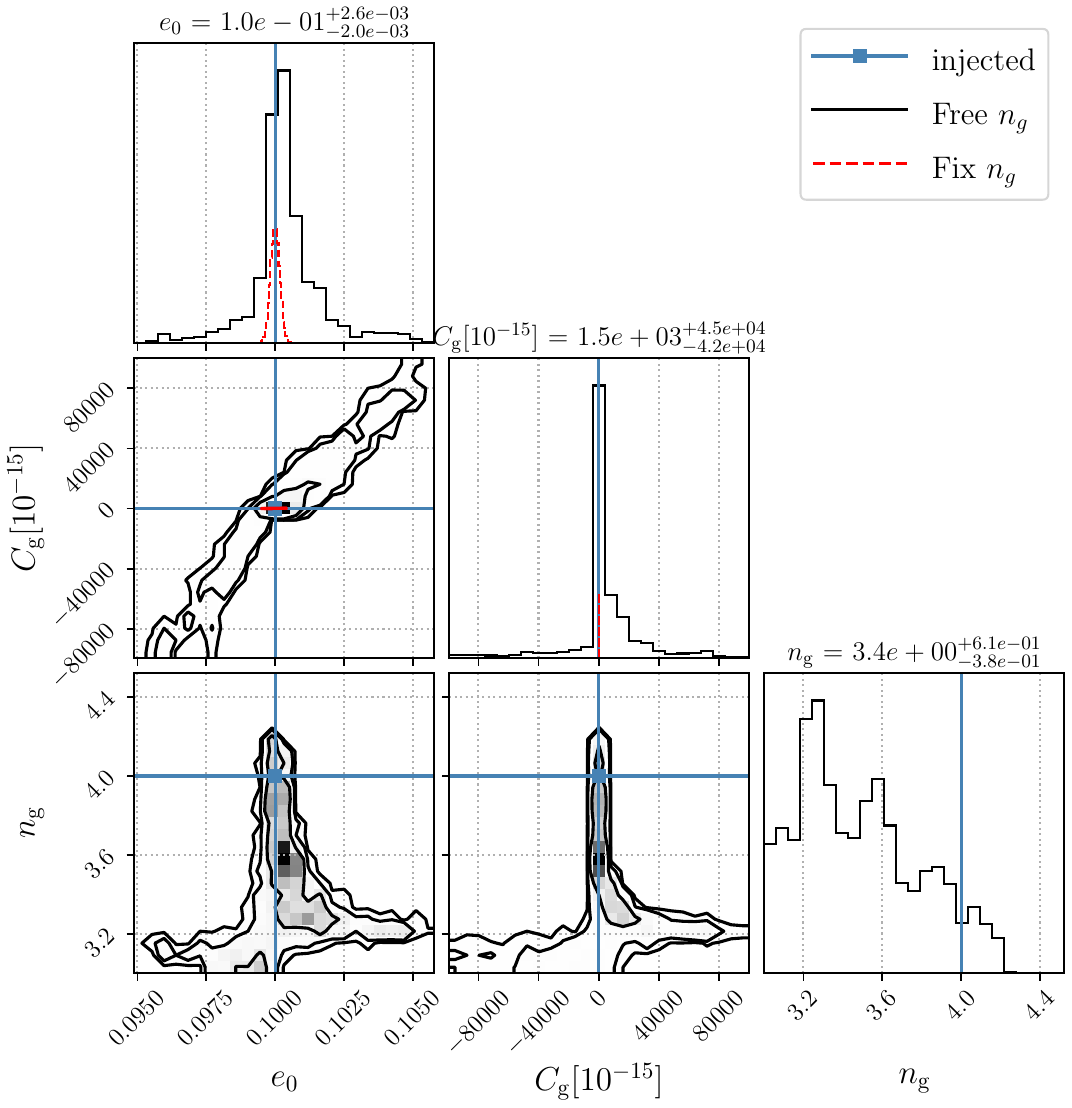}
    \caption{Posteriors for $e_0$, $C_{\rm g}$, and $n_{\rm g}$, when the environmental power-law is a free parameter (black solid). Posteriors for intrinsic-merger parameters are almost the same as in Fig.~\ref{fig:MCMC}. For comparison, we also show $e_0$ and $C_{\rm g}$ posteriors for our fiducial case where $n_{\rm g}$ is kept fixed (dashed red).}
    \label{fig:MCMC_4yr_ng}
\end{figure}

Next, in Fig.~\ref{fig:MCMC_4yr_ng}, we show posteriors for the intrinsic-inspiral parameters when the environmental power-law exponent ($n_{\rm g}$) is also a free variable together with all the other intrinsic parameters. We find that keeping $n_{\rm g}$ free leads to non-Gaussian posteriors for $e_0$, $C_{\rm g}$, and $n_{\rm g}$, and causes apparent biases in the 1D marginal posteriors for both the power-law exponent and the environmental amplitude. We are using zero noise injections which means that the maximum value of the log-likelihood, i.e., zero, is at the injected parameters. However, this does not mean that there can not be other local maxima in the posterior. There is a degeneracy between $C_{\rm g}$ and $n_{\rm g}$ in the environmental phase correction $\Delta\psi_{\rm gas}$ in Eq.~\ref{eq:phasegas2}. This degeneracy occupies a much larger prior volume around $n_{\rm g}\sim3.2$ in Fig.~\ref{fig:MCMC_4yr_ng} than the true peak of the likelihood, which means that even if the likelihood there is lower than at the injected values, the total weight in the evidence could be comparable to that of the true peak. This would mean that the marginal distributions could favour the secondary mode, which is what we are seeing here. Moreover, Fig.~\ref{fig:MCMC_4yr_ng} implies that a small change in the initial eccentricity can allow a particular value of $n_{\rm g}$ to absorb a large environmental amplitude. 

The low SNR in the early inspiral of the signal, where gas dephasing is significant, coupled with the fact that a small variation in the other intrinsic parameters can compensate for relatively larger variations in $C_{\rm g}$ and $n_{\rm g}$, leads to this behavior. We further illustrate this degeneracy in Appendix~\ref{AppC}. If we bring the source to a redshift $z=0.01$, rather than $z=1$, in order to increase the SNR, then all posteriors become Gaussian as shown in Fig.~\ref{fig:MCMC_z}, illustrating that the low SNR in the inspiral phase of the signal is one of the issues. Future high-resolution hydrodynamical simulations of binary-disc interaction with more physics would allow for the identification of richer features and higher-order environmental phase terms, which may help to break this degeneracy at lower SNRs.

\begin{figure}
    \centering
    \includegraphics[width=0.5\textwidth]{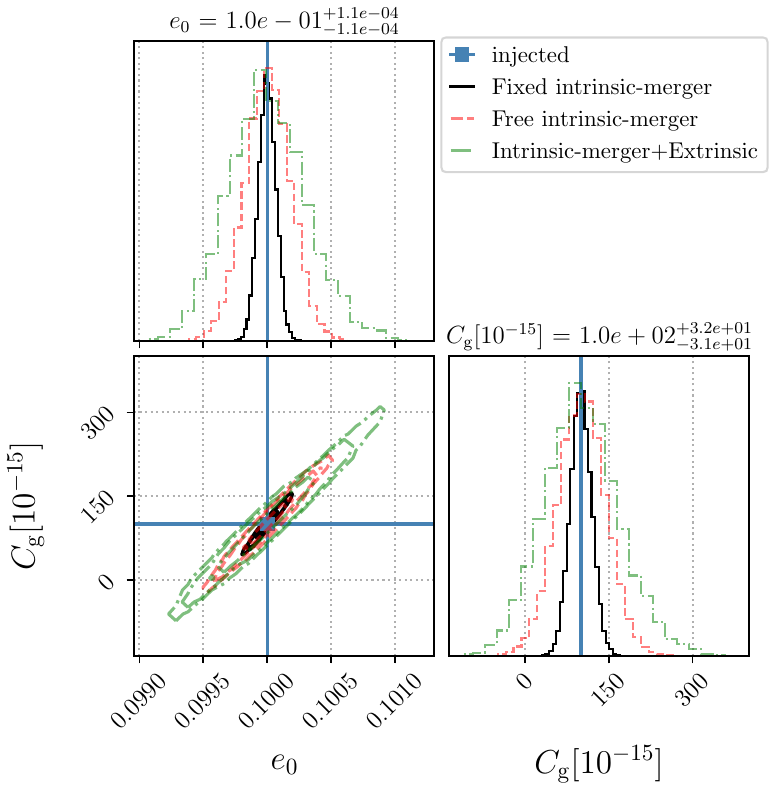}
    \caption{$C_{\rm g}$ and $e_0$ posteriors' comparison for three cases: fixed intrinsic-merger parameters (solid black), free intrinsic-merger parameters (our fiducial case; dashed red) or varying all parameters (dot-dashed green) in Table~\ref{table:parameters} except $n_{\rm g}$.}
    \label{fig:Comp_e0p1_Cg100}
\end{figure}

Last, we compare $e_0$ and $C_{\rm g}$ posteriors between fixed intrinsic-merger parameters, free intrinsic-merger parameters, and free intrinsic-merger and extrinsic parameters in Fig.~\ref{fig:Comp_e0p1_Cg100}.\footnote{See Fig.~E2 of \citet{Garg2024} for posteriors of the extrinsic parameters.} We find that all posteriors peak around the injected values with smallest uncertainties for the fixed intrinsic-merger variables and largest uncertainties for the free extrinsic variables. These results can be attributed to having fewer or more free parameters. For fixed intrinsic-merger parameters, we have no support for $C_{\rm g}=0$. Also, the errors on the intrinsic parameters are almost independent of the inclusion or not of extrinsic parameters in the model. The case with fixed intrinsic-merger parameters arises when we have independent information on those parameters from either merger-ringdown of the same GW signal or EM counterparts. In reality, we will have narrow priors instead of fixed values for $\{M_z,q,\chi_{1,2},t_c\}$ and posteriors on $e_0$ and $C_{\rm g}$ will be somewhere between the black lines and the red dashed lines in  Fig.~\ref{fig:Comp_e0p1_Cg100}.

In the next section, we compute biases and Bayes factors to compare different templates when fitting a given data.

\section{Fitting a wrong template}\label{Sec:Bias}
In this section, we examine the consequences of fitting a wrong template to an injected signal. We compute Bayes factors of the correct template with respect to the wrong template. We always set $\{M_z,q,\chi_1,\chi_2,t_c\}$ to their fiducial values.   

We evaluate Bayes factors by taking the ratio between the evidence ($Z$) of fitting a true template and a false template to a given signal:
\begin{align}
    \mathcal{B}=\frac{Z_{\rm true}}{Z_{\rm false}}.
\end{align}
To reduce statistical variance during PTMCMC, we take the average of Bayes factors from two sets of independent runs, while we report the recovered parameter from the first set of runs. Moreover, we report errors on the Bayes factors by dividing the absolute difference between the Bayes factors from the independent runs by two. The estimated errors on $\ln\mathcal{B}$ suggest small variance between different Bayesian runs.

\subsection{Injecting either only eccentricity or only gas perturbation}
We fit an eccentric template to a gas-perturbed circular injected signal to recover $e_0$, and a gas-perturbed circular template (fixed $n_{\rm g}$) to an injected eccentric signal to recover $C_{\rm g}$. 

In Table~\ref{table:BayesFactor_ecc}, we show the recovered values of $e_0$ and Bayes factors when analysing an injected environmental circular signal for positive and negative values of $C_{\rm g}$ with an eccentric template. The stronger the negative $C_{\rm g}$, the higher the recovered eccentricity from its peaked Gaussian posterior. There are also larger biases on the intrinsic-merger parameters (see Appendix~\ref{AppE}). Since eccentricity can not explain an effect that slows down the inspiral caused by positive $C_{\rm g}$ (an outward torque), it is restricting $e_0$ to small values $\lesssim10^{-3.5}$ with again large biases and one-sided Gaussian posteriors for eccentricity. For weak environmental effects, $C_{\rm g}[10^{-15}]\in\{-10^2,-10^1,10^1\}$, the Bayes factors are inconclusive. However, for strong environmental effects we can confidently accept the true template. Therefore, an EM counterpart or a population-based inference \citep{Garg2024b} will be crucial to be sure about the presence of a weak environmental effect. Furthermore, errors on Bayes factor do not change the conclusion about their decisiveness.

\begin{table}
\centering
    \begin{tabular}{|C|C|C|c|}
        \hline
        \pmb{C_{g,\rm inj}[10^{-15}]} & \pmb{{\color{black}\log_{10}e_0}}&\pmb{{\color{black}\ln\mathcal{B}}}&\textbf{Strength}\\
        \hline
        \hline
        -10^3&-1.6&\cellcolor{green}42.3^{+1.5}_{-1.5}&Decisive\\
        \hline
        -10^2&-2.0&\cellcolor{lime}3.5^{+0.7}_{-0.7}&Inconclusive\\
        \hline
        -10^1&-2.5&\cellcolor{pink}-2.3^{+0.2}_{-0.2}&Inconclusive\\
        \hline
        
        10^1&-4.4&\cellcolor{lime}1.1^{+0.1}_{-0.1}&Inconclusive\\
        \hline
        10^2&-3.8&\cellcolor{green}80.8^{+0.2}_{-0.2}&Decisive\\
        \hline
        10^3&-4.0&\cellcolor{green}310.8^{+0.3}_{-0.3}&Decisive\\
        \hline
    \end{tabular}
\caption{The injected environmental amplitude $C_{g,\rm inj}$, recovered $e_0$, Bayes factor $\ln\mathcal B$ with errors, and the strength of evidence for the true model when fitting an eccentric vacuum model to a circular environmental signal. Here $\ln\mathcal B\gtrsim5$ (green) represents definitive evidence in favor of the true model, $1\lesssim\ln\mathcal B\lesssim3$ (lime) means the true model is weakly preferred, and $-3\lesssim\ln\mathcal B\lesssim-1$ (pink) implies the false model is weakly preferred \citep{Taylor2021}. We term $-3\lesssim\ln\mathcal B\lesssim3$ as inconclusive and $\ln\mathcal B\gtrsim5$ as decisive. We recover negligible eccentricities for positive migration ($C_{\rm g}>0$) since any significant $e_0$ cannot explain slower inspiral.}
\label{table:BayesFactor_ecc}
\end{table}

In Table~\ref{table:BayesFactor_env}, we show the recovered value of $C_{\rm g}$ and Bayes factors, when fitting a gas perturbed circular template to an injected eccentric signal for various small eccentricities. We get narrow-peaked Gaussian posteriors for $C_{\rm g}$. Even for $e_{0,\rm inj}=10^{-2.5}$, we find Bayes factor in favor of the true model, which becomes stronger for higher eccentricity.

\begin{table}
\centering
    \begin{tabular}{|C|C|C|c|}
        \hline
        \pmb{\log_{10}e_{0,\rm inj}} & \pmb{{\color{black}C_{\rm g}}[10^{-15}]}&\pmb{{\color{black}\ln\mathcal{B}}}&\textbf{Strength}\\
        \hline
        \hline  
        -2.5&-12.8&\cellcolor{lime}3.3^{+0.2}_{-0.2}&Inconclusive\\
        \hline
        -2.0&-121.1&\cellcolor{green}5.3^{+0.2}_{-0.2}&Decisive\\
        \hline
        -1.5&-1441.9&\cellcolor{green}58.0^{+0.1}_{-0.1}&Decisive\\
        \hline
    \end{tabular}
\caption{Results obtained when analysing an injected eccentric vacuum signal with a circular environmental template. The columns show the injected eccentricity, $e_{0,\rm inj}$, recovered $C_{\rm g}$, Bayes factor $\ln\mathcal B$, and the strength of evidence in favour of the true model.}
\label{table:BayesFactor_env}
\end{table}

\subsection{Injecting both eccentricity and gas perturbation}
In this section, we fit either only a vacuum eccentric or an environmental circular template to a gas-perturbed eccentric injected signal. Therefore, recovering $e_0$ in the former case and $C_{\rm g}$ in the latter case, respectively.

In Table~\ref{table:BayesFactor_ecc2}, we show the recovered value of $e_0$ when fitting a vacuum eccentric signal to an injected eccentric environmental signal. The posteriors for $e_0$ are narrow-peaked Gaussians. We find that for a strong environmental perturbation ($|C_{\rm g}|=10^{-12}$) PTMCMC heavily favors the true model, while the results are inconclusive for ($|C_{\rm g}|=10^{-13}$). This again emphasizes the need to have a complementary EM signal to have definite proof of an environment.

\begin{table}
\centering
    \begin{tabular}{|C|C|C|C|c|}
        \hline
        \pmb{e_{0,\rm inj}} & \pmb{C_{g,\rm inj}[10^{-15}]} & \pmb{\log_{10}e_0}&\pmb{{\color{black}\ln\mathcal{B}}}&\textbf{Strength}\\
        \hline
        \hline
        10^{-2}&-10^3&-1.6&\cellcolor{green}40.6^{+0.3}_{-0.3}&Decisive\\
        \hline10^{-2}&-10^2&-1.9&\cellcolor{pink}1.2^{+1.2}_{-1.2}&Inconclusive\\
        \hline
        
        10^{-2}&10^2&{-2.4}&\cellcolor{pink}-1.2^{+0.6}_{-0.6}&Inconclusive\\
        \hline
       10^{-2}&10^3&{-2.3}&\cellcolor{green}238.0^{+0.3}_{-0.3}&Decisive\\
        \hline
    \end{tabular}
\caption{The injected $e_{0,\rm inj}$ and $C_{g,\rm inj}$, recovered $e_0$, Bayes factor, and the strength of evidence in favour of the true model, when fitting an injected eccentric environmental signal with a vacuum eccentric template.}
\label{table:BayesFactor_ecc2}
\end{table}

In Table~\ref{table:BayesFactor_env2}, we show the recovered value of $C_{\rm g}$ when studying an injected eccentric environmental signal with a gas-perturbed circular template and find that while the true model is preferred over the false model, the Bayes factors are not decisive.

\begin{table}
\centering
    \begin{tabular}{|C|C|C|C|c|}
        \hline
        \pmb{e_{0,\rm inj}} & \pmb{C_{g,\rm inj}[10^{-15}]} & \pmb{{\color{black}C_{\rm g}}[10^{-15}]}&\pmb{\ln\mathcal{B}}&\textbf{Strength}\\
        \hline
        \hline
        10^{-2}&-10^3&-1122.6&\cellcolor{lime}2.6^{+0.7}_{-0.7}&Inconclusive\\
        \hline
        10^{-2}&-10^2&-220.5&\cellcolor{lime}1.7^{+1.1}_{-1.1}&Inconclusive\\
        \hline
        
        10^{-2}&10^2&-19.9&\cellcolor{lime}2.3^{+0.8}_{-0.8}&Inconclusive\\
        \hline
        10^{-2}&10^3&881.2&\cellcolor{lime}1.6^{+0.6}_{-0.6}&Inconclusive\\
        \hline
    \end{tabular}

\caption{Results obtained when analysing an injected eccentric environmental signal with a circular environmental template. The columns show the injected $e_{0,\rm inj}$ and $C_{g, \rm inj}$, recovered $C_{\rm g}$, Bayes factors $\ln\mathcal B$, and the strength of evidence in favour of the true model.}
\label{table:BayesFactor_env2}
\end{table}

\section{Discussion}\label{Sec:discussion}

The LISA Red Book \citep{Colpi2024}, which reflects the current science objectives of the community, does not consider eccentricity and gas effects in the analysis of future MBHB data. However, recent suites of high-resolution hydrodynamical simulations that embed an eccentric MBHB in a thin accretion disc \citep{Zrake2021,DOrazio2021,Siwek2023,Tiede2024} suggest measurable eccentricities $e_{\rm LISA}\gtrsim 10^{-2.75}$ \citep{Garg2024} in the LISA band despite partial circularization due to GW emission \citep{Peters1964}. These studies also emphasize that a gaseous environment can non-negligibly alter the inspiral GW waveform of MBHBs via observable gas-induced dephasing \citep{Garg2022,Dittmann2023}. Therefore, neglecting eccentricity and gas imprints in the GW waveform could induce bias as shown in Appendix~\ref{AppE}. This work can help to motivate the community to consider both gas and eccentricity as essential parameters for future data analysis.

Usually gas perturbations for MBHBs are measured in terms of a cumulative phase shift in the GW phase using Newtonian waveforms \citep{Garg2022,Dittmann2023}. While this is a good first step to estimate if gas could leave an observable imprint, it makes it impossible to relate this phase shift unequivocally to the presence of a CBD disc. This is because either higher PN corrections, a small eccentricity, or other environmental influences such as a dark matter spike or third body interaction can mimic this effect \citep{Zwick2023}. To be confident that this phase shift is most likely from gas, we need to measure both its effective amplitude ($C_{\rm g}$) and its power-law slope ($n_{\rm g}$). Since gas, via either migration or accretion, induces a phase correction at the $-4$PN order, we not only need high SNR but also numerous cycles in the inspiral phase to detect it. This is why EMRIs are traditionally preferred for this kind of study (e.g., by \citealt{speri2023}), as a $q\gtrsim10^4$ circular system spends $\sim10^5$ GW cycles in band during a four-year LISA observation window, which allows for constraints on both $C_{\rm g}$ and $n_{\rm g}$ even for a moderate Eddington ratio of ${\rm f}_{\rm Edd}=0.1$. However, what MBHBs lack in the number of cycles, they make up some of it by having high SNRs to allow us to measure disc properties within reasonable limits.

Estimating the strength of a gas torque ($\xi$) on the MBHB near-merger is challenging, since most simulations study the system in the regime where GWs are not dominant. Current simulations for non-inspiraling MBHBs predict $\xi\lesssim1$ for a moderately thin disc. There have been studies of the effects of gas on the GW phase of an EMRI in the LISA band \citep{Derdzinski2019,Derdzinski2021,Nouri2023}, albeit using Newtonian-order approximations, but nothing yet for measuring gas torques on coalescing MBHBs aside from \citet{Dittmann2023}.\footnote{There are interesting studies on the relativistic accretion flows onto merging MBHBs which show complex accretion flows, however these are typically not evolved for long enough to measure accurate torques (see e.g. \citealt{Noble2023,Ennoggi2023,Gutierrez2022}).} Many of the simulations with MBHBs that focus on the long-term evolution of the binary neglect magnetic fields and radiative transfer, further increasing the uncertainty in the estimated torques on realistic systems. Therefore, it is not clear whether $\xi$ becomes stronger, weaker, or remains the same as a binary approaches merger. For this reason, in this study we have remained agnostic about the torque strength, and considered a wide-range of $\xi$ which allows for a $|C_{\rm g}|$ as high as $10^{-12}$, in the case that future studies find stronger gas torques in the LISA regime.  

Our parameterization in Eq.~\eqref{eq:Cg} implies that a super-Eddington accretion rate may be responsible for a high value of $C_{\rm g}$ \emph{as well as} a nonlinear dependence of the torque strength $\xi$ in the regime of extremely thin accretion discs, which remains poorly explored. In other words, a degeneracy exists between the disc parameters (in addition to ${\rm f}_{\rm Edd}$ and radiative efficiency) and the resulting torque ($\xi$). This can be broken by future simulations of these systems that explore more representative parameters for luminous AGN systems. Furthermore, inspiraling MBHBs are naturally expected to produce bright EM counterparts, which can provide valuable constraints on ${\rm f}_{\rm Edd}$ and binary eccentricity. The presence of gas allows for a characteristic X-ray emission during the inspiral \citep[see, e.g.][]{Haiman2017,DalCanton2019,Mangiagli2022,Cocchiararo2024}. Detection of such counterparts will not only confirm the presence of an accretion disc, but also provide a narrower prior on the disc parameters from an independent channel. The observation of an EM counterpart will trigger the search for environmental deviations and justify the assumption of fixing $n_{\rm g}=4$, thus allowing us to break the degeneracies between disc parameters such as ${\rm f}_{\rm Edd}$, $\xi$, and the radiative efficiency. The combination of this knowledge with measurements of environmental parameters from the GW signal will provide the strongest constraints on accretion disc structure and binary-disc interaction. 

Section~\ref{Sec:Bias} emphasizes the importance of finding an EM counterpart or performing a population-based inference for moderate $C_{\rm g}$ to conclusively determine that an accretion disc is present. Otherwise, the parameters estimated from the inspiral part of the phase will include biases which can not be completely mitigated by information from the merger-ringdown part of the GW signal. A wrong analysis of the inspiral phase could thus lead to a false detection of eccentricity, and it could even possibly mimic a deviation from GR \citep{Gair2013}. This can have far-reaching consequences, such as raising doubts on the validity of GR, or leading to a biased distribution of MBHB parameters, which would be used to disentangle MBH formation and growth channels. 

There are a number of caveats in the current work. The \textsc{TaylorF2Ecc} model does not include spin-eccentricity cross-terms in the phase, which should be negligible for the small eccentricities we study here. Moreover, we only consider the Newtonian GW amplitude without eccentric and gas-induced corrections. However, the inclusion of higher PN orders and eccentric-environmental cross-terms should only help to improve the measurements of the parameters. While the accuracy of \textsc{TaylorF2Ecc} reduces towards our cutoff at the innermost stable circular orbit, we expect the results will not change if an earlier cutoff is used \citep{Garg2024}. From the astrophysical perspective, the phase correction due to the gas perturbation given by Eq.~\eqref{eq:phasegas2} is a relatively simple model. It ignores any gas torque fluctuations during the binary orbit \citep{Zwick2022} that, in the very early inspiral phase, produce a secular phase shift that mimics our model and can deviate from our fiducial CBD torque (=$\xi\dot{M}\Omega r^2$) even if torque fluctuations themselves are negligible after orbital average. Moreover, once the orbital averaged value of $\dot a_{\rm gas}$ does not well-approximate the semi-major axis decay rate towards the merger, these fluctuations can not be ignored. However, this time-domain effect is cumbersome to include consistently in the frequency-domain waveforms considered here, and we leave this to future work. Also, including higher PN terms in the gas phase would require even higher resolution hydrodynamical simulations so that we could go beyond the Newtonian equations of motion, whose rich features may help us find additional terms. Finally, we have always kept the radiative efficiency $\epsilon$ as $0.1$ even for our highly spinning MBHs. Even if the dependence of gas torque  on $\epsilon$ remains the same near merger, it can vary for the same spin magnitude between prograde or retrograde. We have kept it fixed, and assume that any uncertainty is folded into $\xi$.

\section{Conclusion}\label{Sec:conclusion}
In this paper, we considered GWs from eccentric MBHBs embedded in a CBD to estimate if both eccentricity and gas imprints (migration and accretion) could be concurrently measured from the emitted gravitational waves observed by LISA. We study systems of interest at redshift $z=1$ with highly spinning ($\chi_{1,2}=0.9$) individual BH masses, $M_z$, between $10^4$--$10^6~\MSun$ and a primary-to-secondary mass ratio, $q\in[1.2,8]$, such that the MBHBs spend at least four years in the LISA band before merging. We considered both one-year and four-year times of coalescence ($t_c$) to study the measurability of the parameters for systems with initial eccentricity, $e_0$, from $10^{-3}$--$10^{-1}$ and effective gas amplitude, $C_{\rm g}$, between $-10^{-12}$ and $10^{-12}$. We assumed that the power-law ($n_{\rm g}$) scaling of the environmental effect with semi-major axis is $n_{\rm g}=4$, as expected from both migration and accretion (see Sec.~\ref{Sec:CBD}). Due to the high expected SNRs, $\sim150$--$2500$, we found that LISA observations should be able to place constraints on the intrinsic-inspiral variables, $e_0$ and $C_{\rm g}$, as well as on the intrinsic-merger binary parameters, $\{M_z,q,\chi_{1,2},t_c\}$. To account for LISA's motion around the Sun and model the time delay interferometry response, we used the \textsc{lisabeta} software, and included the dephasing due to gas described by Eq.~\eqref{eq:phasegas2} into the \textsc{TaylorF2Ecc} waveform model. We surveyed the parameter space analytically using the Fisher formalism and then studied a few cases using Bayesian inference. Finally, we assessed whether a weak environmental imprint could be confused with a small eccentric signal using GW data alone. We itemize our main findings below.
\begin{itemize}
    \item Since the gas correction to the GW phase depends linearly on $C_{\rm g}$ in Eq.~\eqref{eq:phasegas2}, the absolute error on $C_{\rm g}$ is independent of its magnitude (see Fig.~\ref{fig:Envamp_ecc0} and Section~\ref{Sec:Fisher1}).
    \item The cross terms between eccentricity and gas are negligible (see Appendix~\ref{AppB}). Therefore, when constraining $C_{\rm g}$ and $e_0$ simultaneously, the relative uncertainties on $e_0$ and $C_{\rm g}$ are independent of their exact magnitude (see Fig.~\ref{fig:Envamp_ecc}).
    \item Using the Fisher formalism, we found that the relative errors on $e_0$ and $C_{\rm g}$ are almost independent of the exact spin magnitude but have the strongest dependence on $M_z$ out of the intrinsic-merger parameters (see Figs~\ref{fig:Envamp_Matrixplot} and \ref{fig:Ecc_Matrixplot}). The constraints on $C_{\rm g}$ and $e_0$ depend upon $q$ and $t_c$ (which set the number of GW cycles) strongly and weakly, respectively. This is due to the rapid versus slow evolution in the very early inspiral of $C_{\rm g}$ and $e_0$, respectively.
    \item $C_{\rm g}\gtrsim10^{-14}$ is constrained to $<50$ per cent relative error for circular binaries and $C_{\rm g}\gtrsim10^{-13}$ for eccentric systems. This translates to confidently measuring ${\rm f}_{\rm Edd}$ to around $0.1$ and $1$ respectively in a four-year observation window for a $M_z=10^5~\MSun$ and $q=8.0$ MBHB embedded in an extremely thin disc and stronger gas torque near-merger (see Fig.~\ref{fig:Envamp_Matrixplot} and the connection between $C_{\rm g}$ and accretion properties in Eq.~\ref{eq:Cg}).
    \item We should be able to measure eccentricities as low as $10^{-2.75}$ in vacuum and as low as $10^{-2}$ in the presence of an accretion disc (see Fig.~\ref{fig:Ecc_Matrixplot}). 
    \item Bayesian inference verifies the results of the Fisher formalism, i.e. posteriors overlapped with the prediction from the Fisher matrix (see Fig.~\ref{fig:MCMC}) and peaked at the same value of $C_{\rm g}$ with or without including eccentricity (see Fig.~\ref{fig:Comp_Cg100}).
    \item Sampling extrinsic parameters does not affect the recovery of $C_{\rm g}$ and $e_0$ (see Fig.~\ref{fig:Comp_e0p1_Cg100}).
    \item Keeping the environmental power-law exponent $n_{\rm g}$ free leads to non-Gaussian and biased posteriors for $\{e_0,C_{\rm g},n_{\rm g}\}$ due to degeneracy between $C_{\rm g}$ and $n_{\rm g}$, and low SNR in the early inspiral (see Fig.~\ref{fig:MCMC_4yr_ng} and Appendix~\ref{AppC}).
    \item An eccentric vacuum template can mimic a circular environmental signal for a weak injected gas amplitude, $|C_{g,\rm inj}|\lesssim10^{-14}$. However, vice versa does not hold for small $e_{0,\rm inj}\gtrsim10^{-2.5}$ (see Tables~\ref{table:BayesFactor_ecc} and \ref{table:BayesFactor_env}).
    \item An injected eccentric environmental signal could be confused with a vacuum eccentric signal for $|C_{g,\rm inj}|\lesssim10^{-13}$ and $e_{0,\rm inj}=0.01$. Similarly, the same signal could be mimicked by a circular gas-perturbed template even for $|C_{g,\rm inj}|\sim10^{-12}$  (see Tables~\ref{table:BayesFactor_ecc2} and \ref{table:BayesFactor_env2}).
    \item A stronger environmental perturbation or a higher eccentricity leads to proportionally large biases on the intrinsic-merger parameters if fitted with a wrong model (see Appendix~\ref{AppE}).
\end{itemize}

\section*{Data availability statement}

The data underlying this article will be shared on reasonable request to the authors.

\section*{Acknowledgements}
MG, AD, and LM acknowledge support from the Swiss National Science Foundation (SNSF) under the grant 200020\_192092. ST is supported by the SNSF Ambizione Grant Number: PZ00P2-202204. We thank the anonymous referee for helpful comments that improved this work. We acknowledge John G. Baker and Sylvain Marsat for providing us access to their \textsc{lisabeta} software. The authors also acknowledge use of the Mathematica software \citep{Mathematica}, NumPy  \citep{harris2020array}, and inspiration drawn from the \textsc{GWFAST} package \citep{Iacovelli2022} regarding the python implementation of \textsc{TaylorF2Ecc}. 

\scalefont{0.94}
\setlength{\bibhang}{1.6em}
\setlength\labelwidth{0.0em}
\bibliographystyle{mnras}
\bibliography{EccEnv}
\normalsize

\appendix
\input{Appendices/AppA}
\input{Appendices/AppB}
\input{Appendices/AppC}
\input{Appendices/AppD}
\input{Appendices/AppE}

\bsp 
\label{lastpage}
\end{document}

%% file: Appendices/AppA.tex
\section{Definition of different terms}\label{AppA}
In Table~\ref{table:Def_var}, we summarize various quantities that are used repeatedly in the main text.
\begin{table}
\centering
    \begin{tabular}{|p{0.3\linewidth}|p{0.6\linewidth}|}
        \hline
        {\rm \bf Terms}&{\bf Definition}\\
        \hline
        \hline
        $\eta$ &Symmetric mass ratio $q/(1+q)^2$\\
        \hline
        $v$ &Characteristic velocity $(GM_z\pi f/c^3)^{\frac13}$\\
        \hline
        $\psi^{(0)}_{\rm TF2}$ & leading-order circular phase contribution $(3/128\eta)v^{-5}$\\
        \hline 
        $\Delta\psi_{\rm gas}$ & leading-order gas-induced phasing correction in Eq.~\eqref{eq:phasegas2}\\
        \hline 
        Intrinsic-merger parameters & $\{M_z,q,\chi_1,\chi_2,t_c\}$\\
        \hline 
        Intrinsic-inspiral parameters & $\{e_0,C_{\rm g},n_{\rm g}\}$\\
        \hline 
        Extrinsic parameters & $\{D_{\rm L},\imath,\phi_c,\lambda,\beta,\psi\}$\\
        \hline
        Fiducial parameters in the LISA frame & $M_z=10^5~\MSun,q=8,\chi_{1,2}=0.9,t_c=4~{\rm years},e_0=0.1,C_{\rm g}=10^{-13},n_{\rm g}=4,D_{\rm L}=6791.3~{\rm Mpc},\{\imath,\phi_c,\lambda,\beta,\psi\}=0.5~{\rm radians}$\\
        \hline 
    \end{tabular}
\caption{Definition of different variables and terms in the main text.}
\label{table:Def_var}
\end{table}

%% file: Appendices/AppB.tex
\section{Gas-eccentricity cross terms?}\label{AppB}

To consider cross-terms between gas and eccentricity, we can modify $\dot{a}_\text{mig}$ in Eq.~\eqref{eq:adotgascir} and add terms up to $\mathcal{O}(e^2)$:
\begin{align}\label{eq:adotgasecc}
    \dot{a}_\text{mig}=&{\mathcal{A}}(1+\mathcal{A}_1e+\mathcal{A}_2e^2)\left(\frac{a}{GM_z/c^2}\right)^{n_{\rm g}}\dot{\bar a}^{(0)}_{\rm GW},
\end{align}
where the dimensionless parameters $\{{\mathcal{A}},\mathcal{A}_1,\mathcal{A}_2,n_{\rm g}\}$ are assumed to be constants for the observation window and this parametrization should be valid in the low-eccentricity limit.

There are only a few high-resolution hydrodynamical studies which have considered both gas and eccentricity in the case of near-equal mass binary systems embedded in a CBD. \citet{Munoz2019,Munoz2020,Zrake2021,DOrazio2021,Siwek2023} study prograde orbits, and \citet{Tiede2024} focus on retrograde orbits. A recent study by \citet{Siwek2023} has focused on the eccentricity evolution of unequal-mass binaries. They all assume that the binary is accreting at the Eddington rate (i.e., ${\rm f}_{\rm Edd}=1$) and with a radiative efficiency $\epsilon=0.1$, and \citet{ShakuraSunyaev1973} viscosity coefficient $\alpha=0.1$. However, the works listed here do not focus on the GW-dominated regime (i.e., $\dot{a}_{\rm GW}>\dot{a}_{\rm gas}$), which adds further uncertainty about how valid their results are in the LISA band. Still, we include their results here as a starting point for understanding binary semi-major axis and eccentricity coupling in gas. The values of $\{\xi,\mathcal{A}_1,\mathcal{A}_2\}$ inferred from simulations are given in Table~\ref{table:adotcoeff}.

\begin{table}
\centering
    \begin{tabular}{|c|c|C|C|C|C|}
        \hline
        \textbf{From} &\textbf{Motion}&\pmb{q}&\pmb{\xi} & \pmb{\mathcal{A}_1} & \pmb{\mathcal{A}_2}\\
        \hline
        \hline
        DD21&pro&1.0&0.31&-13.61&87.07\\
        \hline
        SWH23&pro&1.0&0.22&-22.07&66.76\\
        \hline
        SWH23&pro&2.0&0.16&-24.13&93.01\\
        \hline
        SWH23&pro&10.0&-0.05&-68.09&385.55\\
        \hline
        TD23&retro&1.0&-1.25&0&0\\
        \hline
    \end{tabular}
\caption{Coefficients for $\dot{a}_\text{gas}$ in Eq.~\eqref{eq:adotgasecc}, from \citet{DOrazio2021} (DD21), \citealt{Siwek2023} (SWH23), and \citet{Tiede2024} (TD23) for ${\rm f}_{\rm Edd}=1.0$, $\epsilon=0.1$, and $\alpha=0.1$. These approximations are valid for $e\lesssim0.15$.}
\label{table:adotcoeff}
\end{table}

For the gas-induced migration torque described by the parameterized form in Eq.~\eqref{eq:adotgasecc}, the leading-order phase contribution can be computed assuming $\dot{a}_{\rm GW}=\dot{a}^{(0)}_{\rm GW}$ in Eq.~\eqref{eq:PhaseBD}:
\begin{align}\label{eq:phasemig2}
    \Delta\psi_{\rm mig}&=-\psi^{(0)}_{\rm TF2}20{\mathcal{A}}v^{-2n_{\rm g}}\bigg[\frac{1}{(n_{\rm g}+4)(2n_{\rm g}+5)}+\nonumber\\
    &\frac{\mathcal{A}_1e_0}{(n_{\rm g}+4+\frac{19}{12})(2n_{\rm g}+5+\frac{19}{6})}\left(\frac{v_0}{v}\right)^{\frac{19}{6}}+\nonumber\\
    &\left(\mathcal{A}_2-\frac{157}{12}\right)\frac{e_0^2}{(n_{\rm g}+4+\frac{19}{6})(2n_{\rm g}+5+\frac{19}{3})}\left(\frac{v_0}{v}\right)^{\frac{19}{3}}\bigg].
\end{align}

Note that even if there is no eccentricity-dependent term in $\dot{a}_{\rm mig}$ (i.e. $\mathcal{A}_1=\mathcal{A}_2=0$),  $\Delta\psi_{\rm mig}$ still has an $e_0^2$ dependence due to $\dot{a}_{\rm GW}^2$ in the denominator of Eq.~\eqref{eq:PhaseGas}.

Both $\mathcal{A}_1$ and $\mathcal{A}_2$ are suppressed by powers of $e_0(v_0/v)^{19/6}$ and multiplied by ${\mathcal{A}}$, which is itself extremely small ($\sim \mathcal{O}(10^{-15})$) as per Eq.~\eqref{eq:ampmig}. Therefore, measurements of $\mathcal{A}_1$ and $\mathcal{A}_2$ either requires them to be extremely large, or we need to observe the signal at a much lower frequency than what LISA can measure. Current simulations suggest that $\mathcal{A}_1$ and $\mathcal{A}_2$ are at most $\mathcal{O}(10^2)$ as per Table~\ref{table:adotcoeff}.\footnote{In this work we focus only on the orbital-averaged value of the torque and neglect fluctuations that can have a higher magnitude \citep{Zwick2022}.} Hence, we drop the cross-terms in Eq.~\eqref{eq:phasemig} with $e_0(v_0/v)^{19/6}$ to recover the circular limit in Eq.~\eqref{eq:phasemig}.

%% file: Appendices/AppC.tex
\section{$C_{\rm \lowercase{g}}$-$\lowercase{n}_{\rm \lowercase{g}}$ degeneracy}\label{AppC}

\begin{figure}
    \centering
    \includegraphics[width=0.5\textwidth]{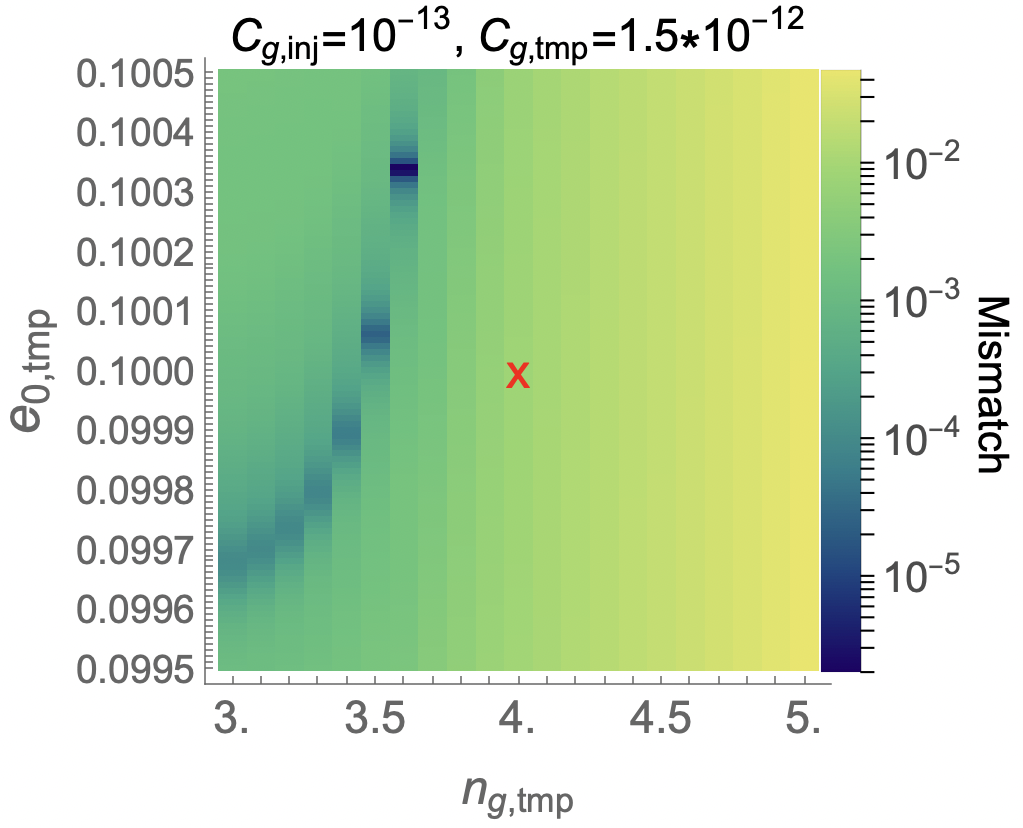}
    \caption{Mismatches between injected and template waveforms as a function of different template eccentricities ($e_{0,\rm tmp}$) and environmental power-laws ($n_{\rm g,tmp}$) with injected intrinsic-inspiral parameters $e_{0,~\rm inj}=0.1$, $C_{\rm g,inj}=10^{-13}$, and $n_{\rm g,inj}=4$ and a template environmental amplitude $C_{\rm g,tmp}=1.5\times10^{-12}$. We vary $e_{0,\rm tmp}$ in the range $[0.0995,0.1005]$ and $n_{\rm g,tmp}$ between $[3,5]$, and mark the injected values with a red cross marker.}
    \label{fig:Match}
\end{figure}

In Fig.~\ref{fig:MCMC_4yr_ng}, we saw that the marginal posteriors on both environmental amplitude and power-law did not peak at their injected values, despite using a zero noise realization. This is due to a degeneracy between the two environmental parameters, $C_{\rm g}$ and $n_{\rm g}$, coupled with the fact that the SNR at $z=1$ is insufficient to break it. To illustrate these points more clearly, we compute mismatches between two waveforms in Fig.~\ref{fig:Match}: an injected waveform ($h_{\rm inj}$) with $e_{0,~\rm inj}=0.1$, $C_{\rm g,inj}=10^{-13}$, and $n_{\rm g,inj}=4$, and a template waveform ($h_{\rm tmp}$) with $C_{\rm g,tmp}=1.5\times10^{-12}$ and varying  $e_{0,\rm tmp}$ and $n_{\rm g,tmp}$, with the rest of the parameters set to their fiducial values for both waveforms. $C_{\rm g,tmp}$ is the same as the recovered value in Fig.~\ref{fig:MCMC_4yr_ng}.

The minimum SNR for which LISA could distinguish between these waveforms with more than $90$ per cent confidence can be computed based on the following criterion for eight free intrinsic parameters \citep{Baird2013}:
\begin{align}\label{eq:match}
    {\rm SNR}_{\rm min}^2= \frac{6.68}{{\rm Mismatch}(h_{\rm inj},h_{\rm tmp})},
\end{align}

Given that the event SNR for our fiducial parameters is $\sim378$, Eq.~\eqref{eq:match} implies that a mismatch of $\geq10^{-4.33}$ is required to distinguish between the injected and template waveforms at $z=1$. However, we find a few combinations of $e_{0,\rm tmp}$ and $n_{\rm g,tmp}$, where the mismatch is lower than that, especially for $e_{0,\rm tmp}>0.1$ and $n_{\rm g,tmp}<3.5$ in Fig.~\ref{fig:Match}. This implies that the SNR at $z=1$ is not enough to distinguish between the two waveforms for those combinations. As the prior volume occupied by the secondary modes is comparatively larger, the \textsc{PTMCMC} sampler correctly preferred those template values over the injected parameters, which led to the wrong recovered values. This behavior is not there if we have a higher SNR, as shown in Fig.~\ref{fig:MCMC_z}.

%% file: Appendices/AppD.tex
\section{Some interesting posteriors}\label{AppD}
\begin{figure}
    \centering
    \includegraphics[width=0.5\textwidth]{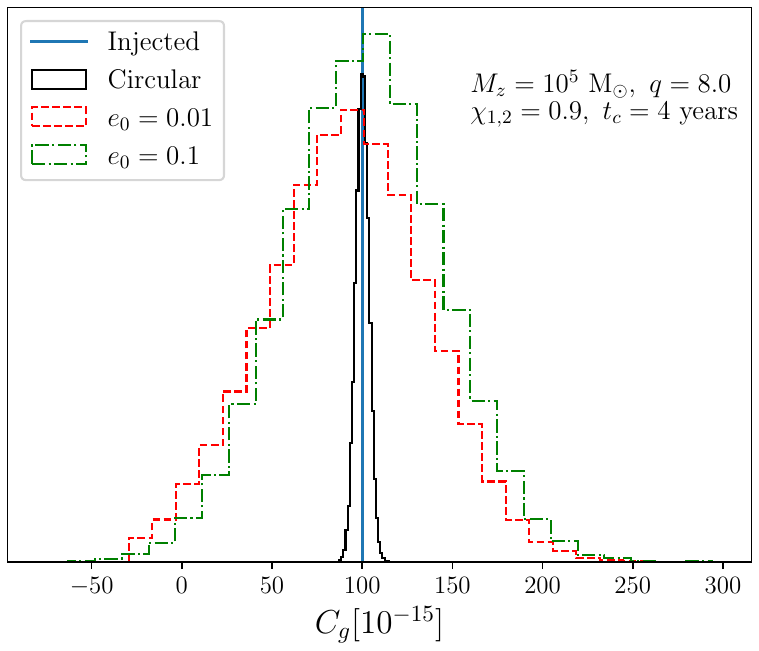}
    \caption{The environmental amplitude posteriors for three eccentricities: $e_0=0$ (solid black),~$e_0=0.01$ (dashed red), and $e_0=0.1$ (dot-dashed green). In all cases, we use the same template to inject the signal and recover it.}
    \label{fig:Comp_Cg100}
\end{figure}

We compare posteriors on the environmental amplitude for a circular system to those for two eccentric systems in Fig.~\ref{fig:Comp_Cg100}. The $C_{\rm g}$ posteriors in all cases peak around the injected value of $10^{-13}$. Also, as expected from Fig.~\ref{fig:Envamp_ecc}, the shape of the posteriors for both non-zero eccentricities are almost the same. In the circular case, we have no support for $C_{\rm g}=0$. The broadening of the posterior is dominated by the extra degree of freedom in the model, rather than by the the presence of eccentricity in the signal itself. 

\begin{figure}
    \centering
    \includegraphics[width=0.5\textwidth]{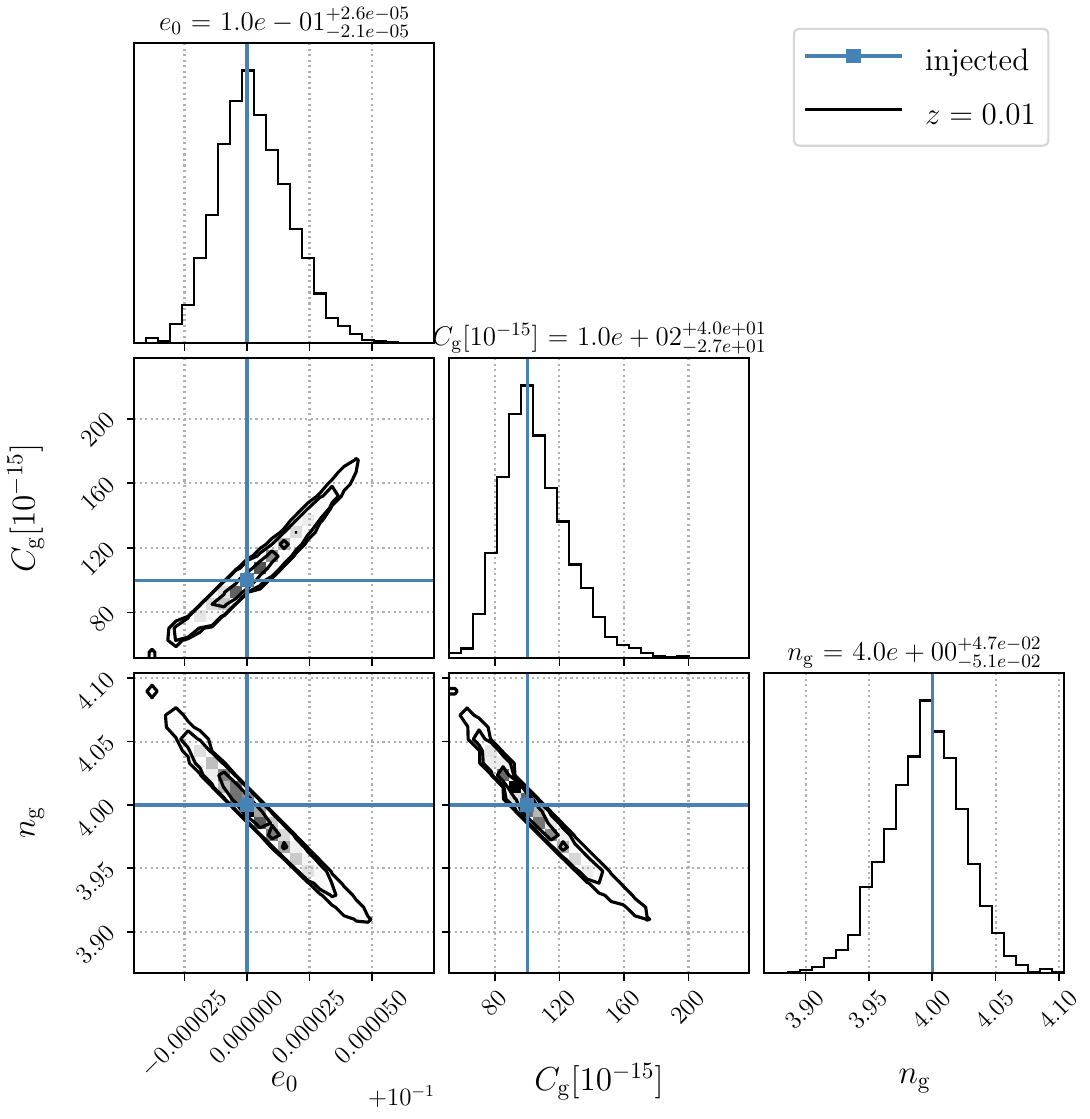}
    \caption{Same as in Fig.~\ref{fig:MCMC_4yr_ng} but at $z=0.01$ instead of $z=1$.}
    \label{fig:MCMC_z}
\end{figure}

In Fig.~\ref{fig:MCMC_z}, we show posteriors for $e_0$, $C_{\rm g}$, and $n_{\rm g}$ at $z=0.01$ to show that high SNR leads to Gaussian posteriors in comparison to non-Gaussian ones in Fig.~\ref{fig:MCMC_4yr_ng},

%% file: Appendices/AppE.tex
\section{Bias due to ignoring gas-perturbation or eccentricity}\label{AppE}
Our goal in this section is to quantify the potential shift in the measured intrinsic-merger parameters (bias) in case that either eccentricity or environment is neglected during analysis of a detected signal. To compute the bias (denoted by $\Delta$) induced on merger parameters due to fitting a wrong template, we take a difference of maximum likelihood values (denoted by $\hat{}$ on top) when recovering an injected signal between a wrong template (denoted by false) and a right template (denoted by true) and divide it by the standard deviation of the given parameter when recovering with the true template.
\begin{align}\label{eq:Bias}
    \Delta\theta&=\frac{\hat{\theta}_{\rm false}-\hat{\theta}_{\rm true}}{\sigma^{\theta}_{\rm true}}.
\end{align}
This way of computation should also minimize statistical uncertainties in the two models during PTMCMC. Moreover, we take average of biases from two sets of independent runs to reduce statistical variance even further. We show results in the following tables~\ref{table:Bias_error_ecc}, ~\ref{table:Bias_error_env}, ~\ref{table:Bias_error_ecc2}, and ~\ref{table:Bias_error_env2} for the same systems of interest that were used in Tables~\ref{table:BayesFactor_ecc}, ~\ref{table:BayesFactor_env}, ~\ref{table:BayesFactor_ecc2}, and ~\ref{table:BayesFactor_env2}, respectively. The stronger that either gas-amplitude or initial eccentricity are, the larger the bias induced in the intrinsic-merger parameters.

\begin{table}
\centering
    \begin{tabular}{|C|C|C|C|C|C|}
        \hline
        \pmb{C_{g,\rm inj}[10^{-15}]}&{\color{black}\pmb{\Delta M_z}}&{\color{black}\pmb{\Delta q}}&{\color{black}\pmb{\Delta \chi_1}}&{\color{black}\pmb{\Delta \chi_2}}&{\color{black}\pmb{\Delta t_c}}\\
        \hline
        \hline
        -10^3&-5.8&-5.9&0.1&1.7&-3.9\\
        \hline
        -10^2&-1.7&-1.8&-0.3&0.8&-1.3\\
        \hline
        -10^1&-0.1&-0.1&0.1&-0.0&0.0\\
        \hline
        10^1&-1.5&-1.5&-0.5&0.8&-1.2\\
        \hline
        10^2&-5.1&-5.2&-1.5&2.7&-4.2\\
        \hline
        10^3&-22.6&-22.8&-9.2&14.7&-20.7\\
        \hline
    \end{tabular}

\caption{The injected environmental amplitude $C_{g,\rm inj}$ and biases on intrinsic-merger parameters due to fitting an vacuum eccentric template to a circular environmental signal.}
\label{table:Bias_error_ecc}
\end{table}

\begin{table}
\centering
    \begin{tabular}{|C|C|C|C|C|C|}
        \hline
       \pmb{\log_{10} e_{0,\rm inj}} &{\color{black}\pmb{\Delta M_z}}&{\color{black}\pmb{\Delta q}}&{\color{black}\pmb{\Delta \chi_1}}&{\color{black}\pmb{\Delta \chi_2}}&{\color{black}\pmb{\Delta t_c}}\\
        \hline
        \hline  
        -2.5&0.0&0.0&0.0&-0.0&0.0\\
        \hline
        -2&2.1&2.1&0.3&-0.9&1.6\\
        \hline
        -1.5&9.2&9.3&0.3&-3.1&6.4\\
        \hline
    \end{tabular}

\caption{The injected eccentricity $e_{0,\rm inj}$ and biases on intrinsic-merger parameters due to fitting a circular environmental template to a vacuum eccentric signal.}
\label{table:Bias_error_env}
\end{table}

\begin{table}
\centering
    \begin{tabular}{|C|C|C|C|C|C|C|}
        \hline
        \pmb{e_{0,\rm inj}} & \pmb{C_{g,\rm inj}[10^{-15}]} &{\color{black}\pmb{\Delta M_z}}&{\color{black}\pmb{\Delta q}}&{\color{black}\pmb{\Delta \chi_1}}&{\color{black}\pmb{\Delta \chi_2}}&{\color{black}\pmb{\Delta t_c}}\\
        \hline
        \hline
        10^{-2}&-10^3&-4.5&-4.5&0.2&1.5&-3.2\\
        \hline
        10^{-2}&-10^2&-1.2&-1.2&-0.3&0.7&-1.0\\
        \hline
        10^{-2}&10^2&1.4&1.4&0.3&-0.8&1.2\\
        \hline
        10^{-2}&10^3&-9.7&-9.8&-2.0&5.2&-8.2\\
        \hline
    \end{tabular}

\caption{The injected eccentricity $e_{0,\rm inj}$ and environmental amplitude $C_{g,\rm inj}$, and biases on intrinsic-merger parameters due to fitting a vacuum eccentric template to an eccentric environmental signal.}
\label{table:Bias_error_ecc2}
\end{table}

\begin{table}
\centering
    \begin{tabular}{|C|C|C|C|C|C|C|}
        \hline
        \pmb{e_{0,\rm inj}} & \pmb{C_{g,\rm inj}[10^{-15}]} &{\color{black}\pmb{\Delta M_z}}&{\color{black}\pmb{\Delta q}}&{\color{black}\pmb{\Delta \chi_1}}&{\color{black}\pmb{\Delta \chi_2}}&{\color{black}\pmb{\Delta t_c}}\\
        \hline
        \hline
        10^{-2}&-10^3&1.8&1.9&0.5&-1.0&1.6\\
        \hline
        10^{-2}&-10^2&1.9&1.9&0.2&-0.9&1.5\\
        \hline
        10^{-2}&10^2&1.7&1.7&0.3&-0.8&1.3\\
        \hline
        10^{-2}&10^3&1.7&1.7&0.1&-0.7&1.3\\
        \hline
    \end{tabular}

\caption{The injected eccentricity $e_{0,\rm inj}$ and environmental amplitude $C_{g,\rm inj}$, and biases on intrinsic-merger parameters due to fitting a circular environmental template to an eccentric environmental signal.}
\label{table:Bias_error_env2}
\end{table}